 \pdfoutput=1
\documentclass[reprint,twocolumn,prx,superscriptaddress]{revtex4-1}
\usepackage{amsmath}
\usepackage{amsfonts}
\usepackage{physics}
\usepackage[caption=false]{subfig}
\usepackage{mathptmx}
\usepackage{bm}
\usepackage{amssymb}
\usepackage{graphicx}

\begin{document}
\title{Driving chemical reactions with polariton condensates}
\author{Sindhana Pannir-Sivajothi}
\affiliation{Department of Chemistry and Biochemistry, University of California
San Diego, La Jolla, California 92093, USA}
\author{Jorge A. Campos-Gonzalez-Angulo}
\affiliation{Department of Chemistry and Biochemistry, University of California
San Diego, La Jolla, California 92093, USA}
\author{Luis A. Mart\'{i}nez-Mart\'{i}nez}
\affiliation{Department of Chemistry and Biochemistry, University of California
San Diego, La Jolla, California 92093, USA}
\author{Shubham Sinha}
\affiliation{Department of Mathematics, University of California San Diego, La
Jolla, California 92093, USA}
\author{Joel Yuen-Zhou}
\email{joelyuen@ucsd.edu}
\affiliation{Department of Chemistry and Biochemistry, University of California
San Diego, La Jolla, California 92093, USA}

\begin{abstract}
When molecular transitions strongly couple to photon modes, they form
hybrid light-matter modes called polaritons. Collective vibrational
strong coupling is a promising avenue for control of chemistry, but
this can be deterred by the large number of quasi-degenerate dark
modes. The macroscopic occupation of a single polariton mode by excitations,
as observed in Bose-Einstein condensation, offers promise for overcoming
this issue. Here we theoretically investigate the effect of vibrational
polariton condensation on the kinetics of electron transfer processes. Compared with excitation with infrared laser sources, the vibrational polariton condensate
changes the reaction yield significantly at room temperature due to additional
channels with reduced activation barriers resulting from the large accumulation of energy in the lower polariton, and the many modes available for
energy redistribution during the reaction. Our results offer tantalizing
opportunities to use condensates for driving chemical reactions, kinetically
bypassing usual constraints of fast intramolecular vibrational redistribution
in condensed phase. 
\end{abstract}
\maketitle

\section*{Introduction}
   Light and matter couple strongly when a large number of molecules are placed within optical cavities that confine light \cite{lidzey1998strong,shalabney2015coherent,long2015coherent}. As a result, hybrid light-matter excitations called polaritons form when a collective molecular transition and a photon mode coherently exchange energy faster than the individual decay from each component. Light-matter strong coupling (SC) opens up a new path to modify material properties by controlling their electromagnetic environment \cite{ebbesen2016hybrid,li2021cavity}. For instance, vibrational strong coupling (VSC), where an infrared
cavity mode couples to an ensemble of localized molecular vibrations in a film or
solution, influences chemical reactivity even without external pumping
\cite{thomas2019tilting,hirai2020modulation}. However,
the microscopic mechanism for modification of molecular processes
through hybridization with light is still poorly understood \cite{galego2019cavity,li2020origin,campos2020polaritonic}, since it could be limited by the presence of a large number of quasi-degenerate
dark modes that do not possess any photonic character and are likely to behave similarly to uncoupled molecules \cite{campos2020polaritonic}.

A Bose-Einstein condensate of polaritons \cite{proukakis2017universal}
offers a solution to this problem since the macroscopic occupation of polaritonic states enhances the effects from SC. In the last decade,
Bose-Einstein condensation has been demonstrated in several organic
exciton-polariton systems at room temperature \cite{daskalakis2014nonlinear,plumhof2014room,dietrich2016exciton,vakevainen2020sub}. Recently, organic polariton condensates were used to build polariton transistors \cite{zasedatelev2019room}, and theoretical predictions suggest they may also modify incoherent charge transport \cite{zeb2020incoherent}. Interestingly, the consequences of polariton condensation on chemical reactivity have not been addressed in the literature prior to the present study.

Ideas of using Bose-Einstein condensates of molecules in chemistry have been previously proposed, but they require ultracold temperatures due to the large mass of the condensing entities \cite{moore2002bose,heinzen2000superchemistry}. The low effective mass that polaritons inherit from their photonic
component, along with the large binding energy of Frenkel excitons,
enables room-temperature condensation \cite{keeling2020bose}. The partly photonic character of polaritons also offers additional benefits such as delocalization and remote action for manipulating chemistry \cite{du2019remote}.

\begin{figure}
\subfloat{ \includegraphics[width=0.4\textwidth]{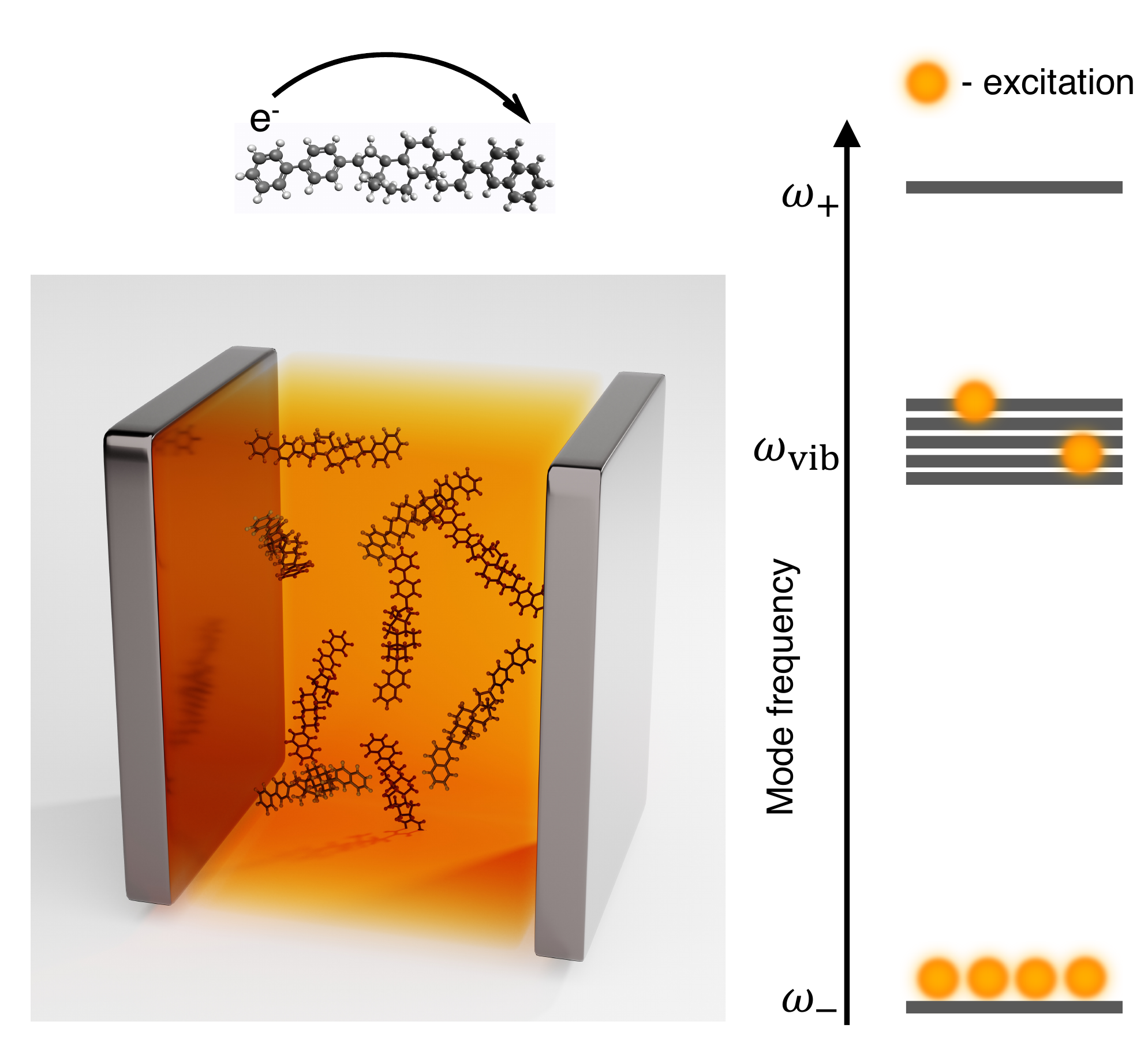}}\caption{\label{fig:schematic}Vibrational polariton condensate. A large number of molecules
are placed inside an optical cavity where their vibrations strongly
couple to the cavity mode. The system is constantly pumped to create
a polariton condensate and the right side of the figure depicts the
occupation of different modes under condensation (frequencies of the
upper polariton, dark modes and lower polariton are $\omega_{+}$,
$\omega_{\text{vib}}$ and $\omega_{-}$, respectively). The rate of intramolecular
electron transfer under vibrational polariton condensation is investigated.}
\end{figure}

Here, we investigate the effect of polariton condensation on electron transfer (ET) (Fig. \ref{fig:schematic}). While the reaction yield under infrared laser excitation,
without SC, already differs from that under thermal equilibrium conditions \cite{delor2014toward,hammes1995vibrationally}, polariton condensation
amplifies this difference by changing the activation
barrier for the forward and backward reactions unevenly, tilting the equilibrium
towards either reactants or products.

\section*{Results}
\subsection*{Bose-Einstein condensation of vibrational polaritons}\label{subsec:bec} 

Bose-Einstein condensation of vibrational polaritons
has not been experimentally achieved yet; however, as we shall argue, there
are compelling reasons to believe that they are presently within reach.
Most theoretical investigations on polariton condensation with organic microcavities involve systems under electronic strong
coupling (ESC) \cite{strashko2018organic,bittner2012estimating};
polariton condensation under VSC requires a separate treatment due to the difference in energy scales and the involved relaxation pathways \cite{del2015quantum}.
While typical bare exciton energies range from 2-3 eV with Rabi splitting
$\sim200$ meV under ESC, the bare frequency of vibrations is 100-300
meV with Rabi splitting $\sim20-40$ meV under VSC. Since the Rabi splitting
is of the order of $k_{\text{B}}T$ at room temperature, thermal effects
are crucial for vibrational polaritons. Under ESC, polariton relaxation
is assisted by high-frequency intramolecular vibrations \cite{somaschi2011ultrafast}, whereas, under VSC, low-frequency solvent modes play a key role in
this process \cite{dunkelberger2016modified,xiang2019state}, similar to what happens in THz phonon Fr{\"o}hlich condensation in biomolecules \cite{frohlich1968bose,zhang2019quantum}.
\begin{figure}
\centering \subfloat{ \includegraphics[width=0.5\textwidth]{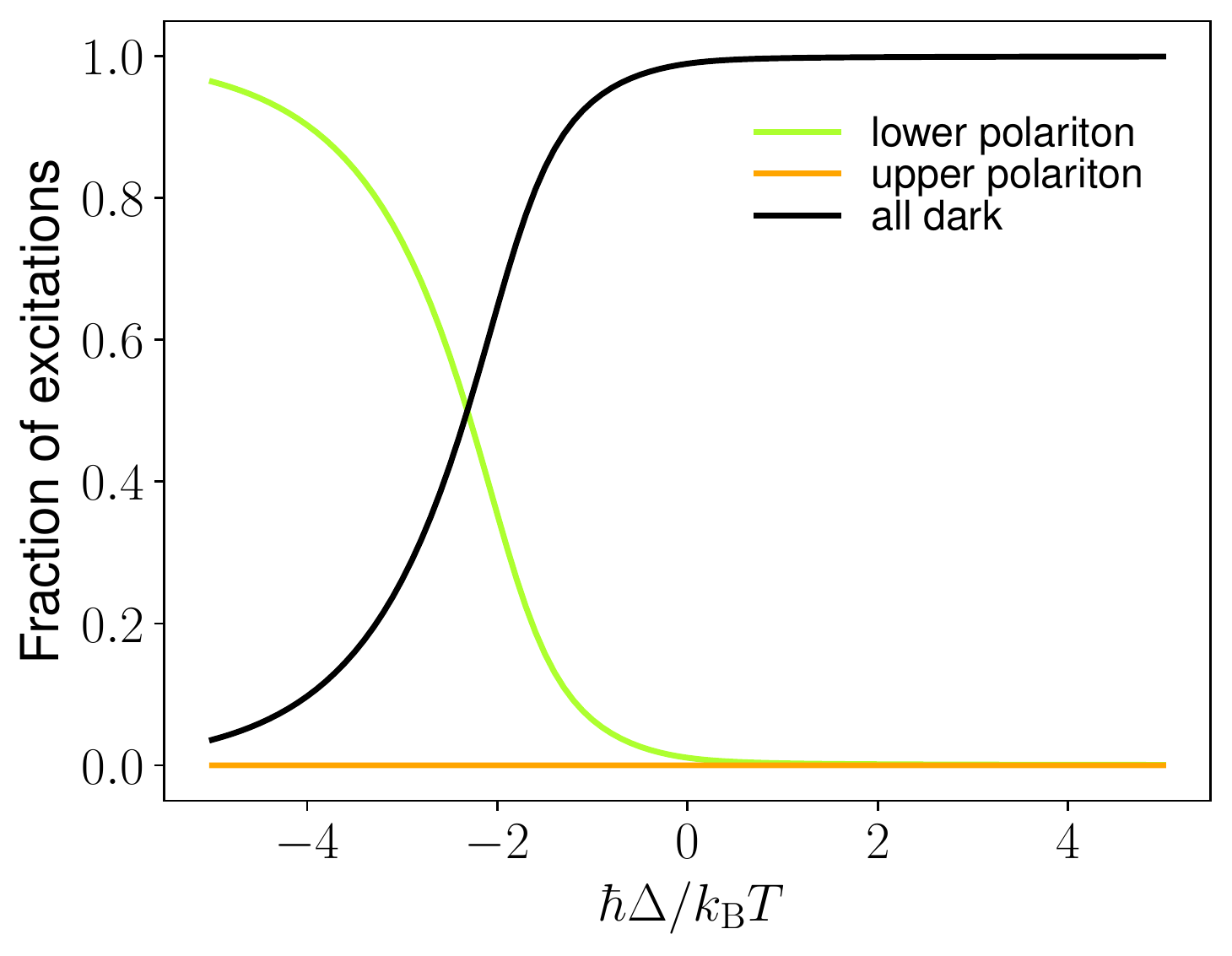}}\caption{\label{fig:Pconst}Polariton condensation transition. Fraction of excitations in different modes as a function of cavity detuning $\hbar\Delta=\hbar\omega_{\text{ph}}-\hbar\omega_{\text{vib}}$ while keeping the pumping rate $P_{-}$ fixed. The black curve includes the excitations in all dark modes taken together while the green and orange curves show the fraction excitations in the lower and upper polariton, respectively. The condensation transition takes place close to $\hbar\Delta\approx-1.5k_{\text{B}}T$. Here, the lower polariton is pumped with rate $P_{-}=0.16N\Gamma_{\downarrow}$, the light-matter coupling $2\hbar g\sqrt{N}=0.72k_{\text{B}}T$, the temperature $k_{\text{B}}T=0.1389\hbar\omega_{\text{vib}}$ ($T=298$K when $\hbar \omega_{\text{vib}}=185$ meV), number of molecules $N=10^7$ and cavity leakage rate $\kappa=\Gamma_{\downarrow}$.}
\end{figure}

We model the polariton system as a set of $N$ vibrational modes ($\hat{a}_{\text{vib},j}$),
with frequency $\omega_{\text{vib}}$, strongly coupled to a single photon
mode ($\hat{a}_{\text{ph}}$) with frequency $\omega_{\text{ph}}$. In the Hamiltonian
of the system, 
\begin{equation}
\begin{aligned}\label{eq:Ham1}
\hat{H}= & \hbar\omega_{\text{ph}}\Big(\hat{a}_{\text{ph}}^{\dagger}\hat{a}_{\text{ph}}+\frac{1}{2}\Big)+\hbar\omega_{\text{vib}}\sum_{j=1}^{N}\Big(\hat{a}_{\text{vib},j}^{\dagger}\hat{a}_{\text{vib},j}+\frac{1}{2}\Big)\\
 & +\sum_{j=1}^{N}\hbar g\Big(\hat{a}_{\text{vib},j}^{\dagger}\hat{a}_{\text{ph}}+\hat{a}_{\text{ph}}^{\dagger}\hat{a}_{\text{vib},j}\Big),
\end{aligned}
\end{equation}
we have applied the rotating wave approximation. Upon diagonalizing this Hamiltonian, we get normal modes: lower and upper polaritons, and $N-1$ dark modes
with frequencies $\omega_{-}$, $\omega_{+}$ and $\omega_{\text{D}}^{k}$,
respectively: 
\begin{equation}
\begin{aligned}\omega_{\pm}= & \omega_{\text{vib}}+\frac{\Delta\pm\Omega}{2},\\
\omega_{\text{D}}^{k}= & \omega_{\text{vib}}\hspace{60pt}2\le k\le N,
\end{aligned}
\label{eq:pol_energy}
\end{equation}
where $\Omega=\sqrt{\Delta^{2}+4g^{2}N}$ is the Rabi splitting and
$\Delta=\omega_{\text{ph}}-\omega_{\text{vib}}$ the detuning between cavity and
molecular vibrations. To model polariton population dynamics, we use Boltzmann rate equations
where the polariton system is weakly coupled to a low-frequency solvent
bath, which enables scattering between modes \cite{banyai2000condensation}.
 These rate equations also account for final-state stimulation, 
\begin{equation}\label{eq:boltz}
\begin{aligned}\frac{dn_{i}}{dt}=\sum_{j}\Big(W_{ij}n_{j}(1+n_{i})-W_{ji}(1+n_{j})n_{i}\Big)-\gamma_{i}n_{i}+P_{i},\end{aligned}
\end{equation}
where $n_{i}$ is the population, $\gamma_{i}$ is the decay rate
and $P_{i}$ is the external pumping rate of the $i^{th}$ mode. The scattering rate from mode $j$ to $i$, $W_{ij}$, satisfies detailed balance: $W_{ij}/W_{ji}=e^{-\beta(\epsilon_{i}-\epsilon_{j})}$. Here, $\beta=1/(k_{\text{B}}T)$, $k_{B}$ is the Boltzmann constant, $T$ is the temperature and $\epsilon_{i}=\hbar \omega_i$ where $\omega_i$ is the frequency of the $i^{th}$ mode. The decay from different modes
is $\gamma_{i}=|c_{\text{vib}}^{i}|^{2}\Gamma_{\downarrow}+|c_{\text{ph}}^{i}|^{2}\kappa$,
where $|c_{\text{vib}}^{i}|^{2}$ and $|c_{\text{ph}}^{i}|^{2}$ are the molecular
and photon fraction, respectively, $\Gamma_{\downarrow}$ is the decay rate of the
molecular vibrations, and $\kappa$ is the cavity leakage rate.

Two factors play a determining role in the condensation threshold: (i) the rate of scattering between polariton and dark modes relative to losses from the system, \textit{i.e.} the rate of thermalization, and (ii) the abundance of modes close in energy to the condensing mode \cite{imamog1996nonequilibrium}. For all calculations, we assume fast thermalization. As mentioned in (ii), the presence of many modes close to the lower polariton would deter
condensation by distributing the energy pumped into the system among
all these modes. Thus, the energetic proximity between the dark state manifold, which has a large density of states (DOS), and the lower polariton poses one of the biggest challenges for polariton condensation under VSC.

The distribution of excitations between the polariton and dark modes is shown in Fig. \ref{fig:Pconst} for different detunings and we observe a condensation transition at $\hbar\Delta\approx -1.5k_{\text{B}}T$ (see Supplementary Note 2 for details). Above the condensation threshold, a large fraction of excitations reside in the lower polariton $\frac{n_{-}}{(\sum_{k=2}^{\infty}n_{\text{D}}^k)}\gg\frac{1}{(N-1)e^{-\beta\hbar(\Omega-\Delta)/2}}$. 

\begin{figure}
\subfloat{\includegraphics[width=0.45\textwidth]{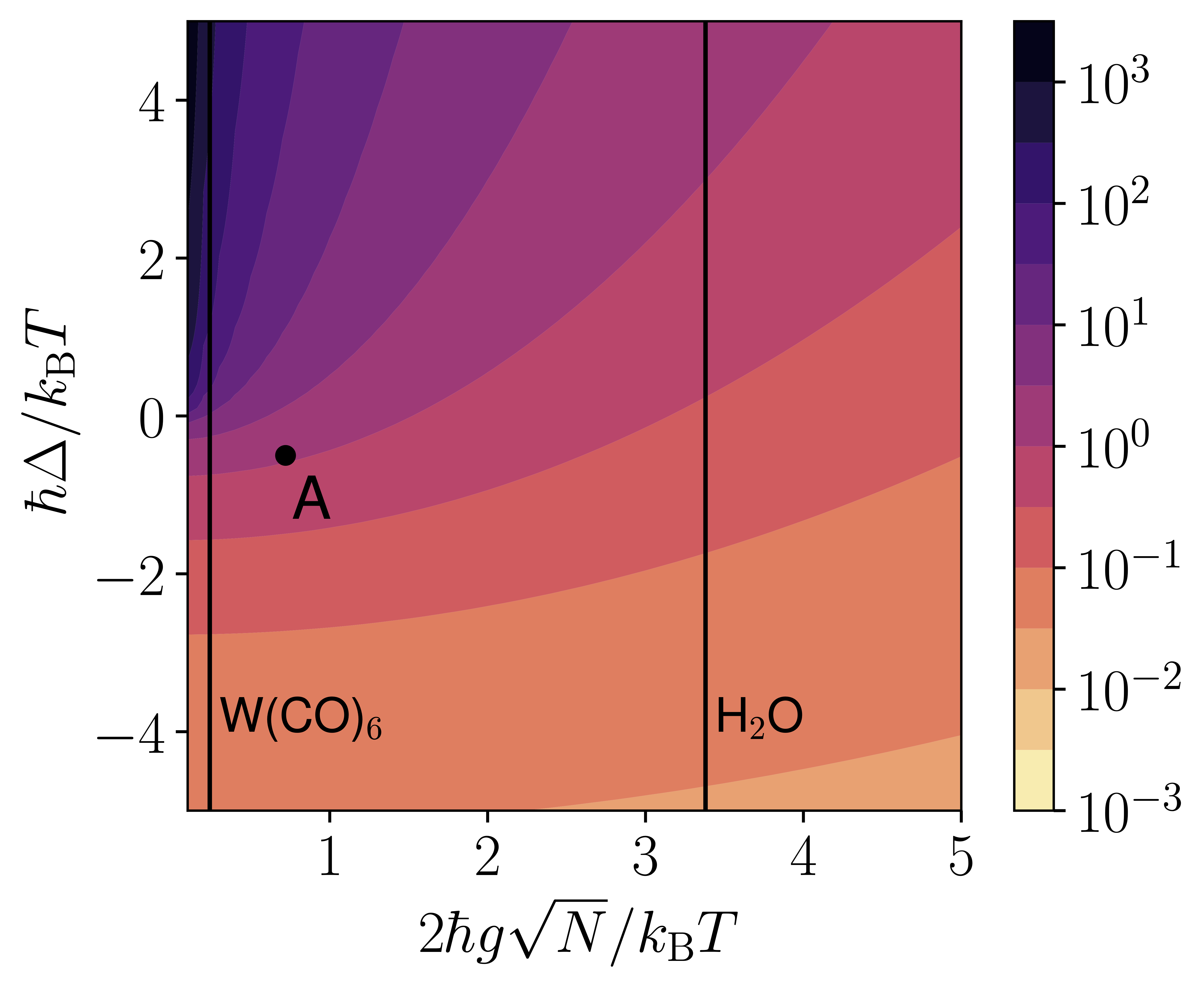}}\caption{\label{fig:Nd}Polariton condensation threshold. Numerically
obtained average population per molecule at the condensation threshold
$\bar{n}=P_{\text{th}}/N\Gamma_{\downarrow}$ ($10\%$ of the excitations are in the lower polariton), for a range of light-matter
coupling strengths $2\hbar g\sqrt{N}$ and cavity detunings $\hbar\Delta=\hbar\omega_{\text{ph}}-\hbar\omega_{\text{vib}}$. In the black and purple regions of the plot ($\Delta>0$ and $2\hbar g\sqrt{N}/k_{\text{B}}T<2$), the threshold for condensation is high, $\bar{n}\gg1$, and polariton condensation is difficult to achieve experimentally. The threshold for condensation is much lower, $\bar{n}<0.1$, for the lighter colored (yellow, orange) regions. In our plot above only the upper polariton is pumped and we use cavity leakage rate $\kappa=\Gamma_{\downarrow}$. The vertical lines correspond to $2\hbar g\sqrt{N}/k_{\text{B}}T$ at room temperature for H$_2$O ($2\hbar g\sqrt{N}\approx 700 \text{cm}^{-1}$) and W(CO)$_6$ ($2\hbar g\sqrt{N}\approx 50 \text{cm}^{-1}$). Calculations in Fig. \ref{fig:yield}-\ref{fig:rateconst} are presented for the conditions in point A. }
\end{figure}

The average population per molecule at the condensation threshold $\bar{n}=P_{\text{th}}/N\Gamma_{\downarrow}$ is a good measure of the feasibility of vibrational polariton condensation. For instance, demanding population inversion, $\bar{n}>0.5$, would be experimentally difficult to achieve in general. In Fig. \ref{fig:Nd}, we plot $\bar{n}$ for different
light-matter coupling strengths, $2\hbar g\sqrt{N}$, and detunings,
$\hbar\Delta$. Here, we numerically obtain $P_{\text{th}}$ as the pumping rate when $10\%$ of the excitations are in the lower polariton. The threshold obtained this way closely corresponds with the theoretical condition for condensation
\begin{equation}
\bar{n}_{\text{D}}^{k}>n_{\text{solvent}}\Big(\frac{\hbar(\Omega-\Delta)}{2}\Big),
\end{equation}
where, $\bar{n}_{\text{D}}^{k}=\frac{1}{N-1}\sum_{k=2}^{N}n_{\text{D}}^{k}$ is the
average occupation of a dark mode, and $n_{\text{solvent}}(E)$ is the Bose-Einstein
population of a solvent mode with energy $E$ at room temperature
$T_{\text{room}}$  \cite{imamog1996nonequilibrium}. The energy difference between the lower polariton and the dark state reservoir $\hbar(\Omega-\Delta)/2$ determines the condensation threshold. From Fig. 3 we see that vibrational polariton condensation is feasible for water even at room temperature for up to zero detuning.

Our model does not include disorder; as a result, all dark modes are degenerate at frequency $\omega_{\text{vib}}$, but in experimental systems, inhomogeneous
broadening of transitions can lead to non-zero density of dark states
even at the bottom of the lower polariton branch \cite{vurgaftman2020negligible}.
This fact will affect the condensation threshold, and should be considered in the future while looking for experimental systems that can demonstrate vibrational
polariton condensation. Stimulating the lower polariton directly by
shining a resonant laser on it \cite{zasedatelev2019room} or using a Raman scattering scheme \cite{del2016exploiting} can help overcome this issue by dynamically
lowering the condensation threshold.
 
\subsection*{Chemical reactions and vibrational polariton condensation} \label{subsec:reaction}

Electron transfer has been theoretically studied under both ESC \cite{herrera2016cavity,semenov2019electron}
and VSC \cite{campos2019resonant,phuc2020controlling}. Here, we look at how vibrational polariton
condensation affects the rate of intramolecular nonadiabatic electron
transfer using the VSC version \cite{campos2019resonant} of the Marcus-Levich-Jortner
(MLJ) model \cite{marcus1964chemical,levich1966present,jortner1976temperature}. 

Our system consists of $N$ molecules placed inside an optical cavity
supporting a single photon mode with bosonic operator $\hat{a}_{\text{ph}}$
and frequency $\omega_{\text{ph}}$. The molecules can be in the reactant
$\text{R}$ or product $\text{P}$ electronic state; for the $i^{th}$ molecule,
these states are denoted by $\ket{\text{R}_{i}}$ and $\ket{\text{P}_{i}}$, respectively.
Each electronic state is dressed with a high-frequency intramolecular vibrational mode with bosonic operator $\hat{a}_{x,i}$ and frequency
$\omega_{\text{vib}}$ where $x=\text{R},\text{P}$; this mode couples to the photon mode. The equilibrium geometry of this vibrational mode depends on the electronic
state according to, 
\begin{equation}
\hat{a}_{\text{R},i}=\hat{D}_{i}^{\dagger}\hat{a}_{\text{P},i}\hat{D}_{i},\label{eq:displace}
\end{equation}
where $\hat{D}_{i}=\exp((\hat{a}_{\text{P},i}^{\dagger}-\hat{a}_{\text{P},i})\sqrt{S})$ and $S$ is the Huang-Rhys factor.

Apart from the intramolecular vibrations, an effective low-frequency
solvent mode surrounding each molecule facilitates ET. It is treated classically, with
$\mathbf{q}_{\text{S},i}$ and $\mathbf{p}_{\text{S},i}$ being its position and
momentum.

The Hamiltonian $\hat{H}$ for the full system is a generalization of Eq. (\ref{eq:Ham1}) to account for the chemical reaction,
\begin{equation}
\hat{H}=\hat{H}_{0}+\hat{V}_{\text{react}},
\end{equation}
and
\begin{equation}
\begin{aligned}\hat{H}_{0}= &\hat{H}_{\text{ph}}+\sum_{i=1}^{N}\sum_{x=\text{R},\text{P}}(\hat{H}_{x,i}+\hat{V}_{x,i})\ket{x_{i}}\bra{x_{i}},\\
\hat{V}_{\text{react}}=&\sum_{i=1}^{N}J_{\text{R}\text{P}}\Big(\ket{\text{R}_{i}}\bra{\text{P}_{i}}+\ket{\text{P}_{i}}\bra{\text{R}_{i}}\Big).
\end{aligned}
\label{eq:Hamil1}
\end{equation}
where $\hat{H}_{0}$ describes the photon ($\hat{H}_{\text{ph}}$), intramolecular vibrations and solvent
modes of the $i^{th}$ molecule ($\hat{H}_{x,i}$), and light-matter couplings ($\hat{V}_{x,i}$). The diabatic
coupling $\hat{V}_{\text{react}}$ is a perturbation that couples $\text{R}$ and
$\text{P}$ electronic states with coupling strength $J_{\text{R}\text{P}}$. We have taken the dipole moment to be zero when the vibrational coordinate is in its equilibrium position in both $\text{R}$ and $\text{P}$ electronic states. Relaxing this assumption will add terms of the form $c_x(\hat{a}_{\text{ph}}+\hat{a}_{\text{ph}})\ket{x_i}\bra{x_i}$ to the Hamiltonian and will not affect the reaction rates calculated.
\begin{equation}
\begin{aligned}\hat{H}_{\text{ph}} & =\hbar\omega_{\text{ph}}\Big(\hat{a}_{\text{ph}}^{\dagger}\hat{a}_{\text{ph}}+\frac{1}{2}\Big),\\
\hat{H}_{\text{R},i} & =\hbar\omega_{\text{vib}}\Big(\hat{a}_{\text{R},i}^{\dagger}\hat{a}_{\text{R},i}+\frac{1}{2}\Big)+\frac{1}{2}\hbar\omega_{\text{S}}\Big(|\mathbf{p}_{\text{S},i}|^{2}+|\mathbf{q}_{\text{S},i}+\mathbf{d}_{\text{S}}|^{2}\Big),\\
\hat{H}_{\text{P},i} & =\hbar\omega_{\text{vib}}\Big(\hat{a}_{\text{P},i}^{\dagger}\hat{a}_{\text{P},i}+\frac{1}{2}\Big)+\frac{1}{2}\hbar\omega_{\text{S}}\Big(|\mathbf{p}_{\text{S},i}|^{2}+|\mathbf{q}_{\text{S},i}|^{2}\Big)+\Delta G,\\
\hat{V}_{x,i} & =\hbar g_{x}(\hat{a}_{x,i}^{\dagger}\hat{a}_{\text{ph}}+\hat{a}_{\text{ph}}^{\dagger}\hat{a}_{x,i}),
\end{aligned}
\label{eq:Hamil2}
\end{equation}
where $\Delta G$ is the free-energy difference of each individual molecule reaction.

We construct potential energy
surfaces (PES) by parametrically diagonalizing $\hat{H}_{0}$ as a function of the solvent coordinate $\mathbf{q}_{\text{S},i}$. The operators $\hat{N}_{\text{R}}=\sum_{i=1}^{N}\ket{\text{R}_{i}}\bra{\text{R}_{i}}$
and $\hat{N}_{\text{P}}=\sum_{i=1}^{N}\ket{\text{P}_{i}}\bra{\text{P}_{i}}$ commute with $H_0$ and correspond to the number of $\text{R}$ and $\text{P}$ molecules, respectively. While dynamics under $\hat{H}_{0}$ conserves $N_{\text{R}},N_{\text{P}}$, the effect of $\hat{V}_{\text{react}}$ is to induce reactive transitions that modify those quantities while keeping $N=N_{\text{R}}+N_{\text{P}}$ constant. We assign the label $1\le i\le N_{\text{R}}$ to $\text{R}$ molecules, and $N_{\text{R}}+1\le i\le N$ to $\text{P}$ molecules. We also reorganize the intramolecular vibrations into a single bright mode, 
\begin{equation}
\hat{a}_{\text{B}(N_{\text{R}},N_{\text{P}})}=\frac{1}{\sqrt{g_{\text{R}}^{2}N_{\text{R}}+g_{\text{P}}^{2}N_{\text{P}}}}\Bigg(g_{\text{R}}\sum_{i=1}^{N_{\text{R}}}\hat{a}_{\text{R},i}+g_{\text{P}}\sum_{i=N_{\text{R}}+1}^{N}\hat{a}_{\text{P},i}\Bigg),\label{eq:bright}
\end{equation}
that possesses the correct symmetry to couple with light and $N-1$
dark modes ($\text{D}_{k}$), 
\begin{equation}
\hat{a}_{\text{D}(N_{\text{R}},N_{\text{P}})}^{k}=\sum_{i=1}^{N_{\text{R}}}c_{k,i}\hat{a}_{\text{R},i}+\sum_{i=N_{\text{R}}+1}^{N}c_{k,i}\hat{a}_{\text{P},i},
\end{equation}
labeled by an additional index $2\le k\le N$, which do not couple
with light. The dark modes are orthogonal to the bright mode $g_{\text{R}}\sum_{i=1}^{N_{\text{R}}}c_{k,i}+g_{\text{P}}\sum_{i=N_{\text{R}}+1}^{N}c_{k,i}=0$,
and to each other $\sum_{i=1}^{N}c_{k,i}c_{k',i}^{*}=\delta_{k,k'}$. Unless mentioned otherwise, the number of $\text{R}$
and $\text{P}$ molecules is $N_{\text{R}}$ and $N_{\text{P}}$, respectively, and for brevity, we will drop $(N_{\text{R}},N_{\text{P}})$ dependence in the subscript. The bright and photon modes combine to
form the upper polariton ($\text{UP}$) $\hat{a}_{+}$, and lower polariton
($\text{LP}$) $\hat{a}_{-}$, modes: 
\begin{equation}
\begin{aligned}\hat{a}_{+} & =\cos\theta\hat{a}_{\text{ph}}+\sin\theta\hat{a}_{\text{B}},\\
\hat{a}_{-} & =\sin\theta\hat{a}_{\text{ph}}-\cos\theta\hat{a}_{\text{B}},
\end{aligned}
\label{eq:polaritons}
\end{equation}
with mixing angle, 
\begin{equation}
\theta=\tan^{-1}\Bigg[\frac{\Omega-\Delta}{2\sqrt{g_{\text{R}}^{2}N_{\text{R}}+g_{\text{P}}^{2}N_{\text{P}}}}\Bigg],\label{eq:UPLPmode}
\end{equation}
where $\Omega=\sqrt{\Delta^{2}+4(g_{\text{R}}^{2}N_{\text{R}}+g_{\text{P}}^{2}N_{\text{P}})}$
is the Rabi splitting, and $\Delta=\omega_{\text{ph}}-\omega_{\text{vib}}$ the
detuning between cavity and molecular vibrations. The eigenstates
of $\hat{H}_{0}$ are the dark, upper and lower polariton modes with
frequencies given in Eq. (\ref{eq:pol_energy}).

According to the MLJ theory, the rate constant for ET outside
of an optical cavity depends on properties of the intramolecular and solvent
modes \cite{marcus1964chemical,levich1966present,jortner1976temperature}.
Under laser driving, this rate constant is, 
\begin{equation}
\label{eq:outside}
\begin{aligned}k_{\text{R}\rightarrow \text{P}}^{\text{IR}} & =\sum_{n=0}^{\infty}P_{\bar{n}}(n)k_{\text{R}\rightarrow \text{P}}(n)\end{aligned}
\end{equation}
where 
\begin{equation}
\begin{aligned}k_{\text{R}\rightarrow \text{P}}(n)= & \sqrt{\frac{\pi}{\lambda_{\text{S}}k_{
				\text{B}}T}}\frac{|J_{\text{R}\text{P}}|^{2}}{\hbar}\sum_{f=-n}^{\infty}|\bra{n}\ket{n+f}'|^{2}\exp(-\frac{E_{f}^{\ddagger}}{k_{\text{B}}T}),\\
P_{\bar{n}}(n)= & e^{-\bar{n}}\frac{\bar{n}^{n}}{n!},\\
E_{f}^{\ddagger}= & \frac{(\Delta G+\lambda_{\text{S}}+f\hbar\omega_{\text{vib}})^{2}}{4\lambda_{\text{S}}},\\
\bra{n}\ket{n+f}'= & \bra{n}\hat{D}_i\ket{n+f}.
\end{aligned}
\end{equation}
Here, $P_{\bar{n}}(n)$ is the Poisson distribution with average mode
population $\bar{n}$, $\lambda_{\text{S}}$ is the solvent reorganization
energy,
$E_{f}^{\ddagger}$ is the activation energy, and $|\bra{n}\ket{n+f}'|^{2}$
is the Franck-Condon (FC) factor, where $\ket{n}$ and $\ket{n+f}'$
are the intramolecular initial and final states, respectively. $P_{\bar{n}}(n)$ has been taken to correspond to the
ideal laser driven-damped harmonic oscillator, leading to a coherent
state in the vibrational mode. The presence of anharmonic
couplings would lead to intramolecular vibrational energy redistribution (IVR) \cite{nesbitt1996vibrational}, reducing the value of $P_{\bar{n}}(n)$ for high-lying
Fock states. However, as we shall see below, even under these ideal
circumstances, the condensate can outcompete the laser-driven situation
in terms of reactivity. We thus expect the benefits of the condensate
to be enhanced when IVR processes are taken into account.

Apart from vibrations within the reacting molecule, under VSC, the ET process also depends on vibrations in all other molecules and the photon mode, and can be represented by,
\begin{equation}
\sum_{k=2}^{N}\text{D}_{k}+\text{LP}+\text{UP}\rightarrow\sum_{k=2}^{N}\text{D}_{k}'+\text{LP}'+\text{UP}'.
\end{equation}
Here and hereafter, the primed and unprimed quantities refer to electronic
states with $(N_{\text{R}},N_{\text{P}})$ and $(N_{\text{R}}-1,N_{\text{P}}+1)$ reactant-product
distributions, respectively. The symmetry of the light-matter coupling
allows us to use the dark state basis introduced in \cite{strashko2016raman}
and \cite{campos2019resonant} to reduce the number of modes involved
in the reaction from $N+1$ to three, 
\begin{equation}
\text{D}_{\text{R},c}+\text{LP}+\text{UP}\rightarrow \text{D}_{\text{P},c}'+\text{LP}'+\text{UP}'.\label{eq:reac_pol}
\end{equation}
Here, the $c^{th}$ molecule is reacting, while $\text{D}_{x,c}$ and $\text{D}_{x,c}'$
are dark modes highly localized in it, with corresponding operators
$\hat{a}_{\text{D}}^{(R,c)}$ and $\hat{a}_{\text{D}}^{(P,c)\prime}$ (see Supplementary
Note 1).

 We perform all our calculations in this subsection using parameters from point A in Fig. \ref{fig:Nd} but while pumping the lower polariton. Here, $\hbar\Delta=-0.5k_{\text{B}}T$, $2\hbar g\sqrt{N}=0.72k_{\text{B}}T$, $k_{\text{B}}T=0.1389\hbar\omega_{\text{vib}}$ ($T=298$K when $\hbar \omega_{\text{vib}}=185$ meV) and $N=10^7$; we choose pumping rate $P_{-}=0.08N\Gamma_{\downarrow}$, which leads to average mode populations $N_+=0.064$, $N_-=1.94 \times 10^4$ and $N_{\text{D}}=0.078$ under symmetric coupling $g_{\text{R}}=g_{\text{P}}=g$. Here, $2.4\%$ of all excitations reside in the lower polariton. To compare the reaction rates under polariton condensation and outside the cavity under pumping, we take $\bar{n}=0.08$ in Eq. (\ref{eq:outside}). Under condensation, the initial vibrational state of the system can be described by $\rho=\sum_{n_{+},n_{-},n_{\text{D}}}P(n_{+},n_{-},n_{\text{D}})\ket{n_{+},n_{-},n_{\text{D}}}\bra{n_{+},n_{-},n_{\text{D}}}$, where the entries in $\ket{n_{+},n_{-},n_{\text{D}}}$ label number of quanta in the $\text{UP}$, $\text{LP}$ and $\text{D}_{\text{R},c}$
 modes, respectively. The results from the rate equations Eq. (\ref{eq:boltz}) provide us only with the average steady-state mode populations, $N_{+}$, $N_{-}$ and $N_{\text{D}}$, and not the distribution $P(n_{+},n_{-},n_{\text{D}})$. For simplicity, we assume the semiclassical approximation $P(n_{+},n_{-},n_{\text{D}})\approx \delta_{n_{+},0}P^{\text{th}}_{N_{\text{D}}}(n_{\text{D}})\delta_{n_{-},N_{-}}$, where $P^{\text{th}}_{N_{\text{D}}}(n)$ is the thermal distribution with average population $N_{\text{D}}$. This approximation is reasonable for populations $N_{+}<N_{\text{D}}\ll1\ll N_{-}$,
\begin{equation}
\rho=\sum_{n_{\text{D}}}P_{N_{\text{D}}}^{\text{th}}(n_{\text{D}})\ket{0,N_{-},n_{\text{D}}}\bra{0,N_{-},n_{\text{D}}}.
\end{equation}
 The product vibrational states are $\ket{\nu_{+},\nu_{-},\nu_{\text{D}}}'$. 

We assume that cavity leakage and rate of scattering between modes
is much faster than the rate of the chemical reaction. For a cavity
with $\sim100$ ps lifetime and ET reactions with $1/k_{\text{R}\rightarrow \text{P}}\sim10^{6}-10^{2}$
ps \cite{miller1984intramolecular}, this assumption is valid. Therefore,
if the populations of polariton modes change during the course
of reaction, they quickly reach a steady state before the next molecule
reacts. Similarly, we also assume that the polariton and dark mode populations reach a steady state before the backward reaction takes place while computing the rate constant $k_{\text{P}\rightarrow \text{R}}^{\text{cond}}$. Generalizing the cavity MLJ theory presented in \cite{campos2019resonant},
we calculate the rate constant 
\begin{equation}
k_{\text{R}\rightarrow \text{P}}^{\text{cond}}=\sum_{n=0}^{\infty}P_{N_{\text{D}}}^{\text{th}}(n)k_{\text{R}\rightarrow \text{P}}^{\text{cond}}(n)\label{eq:rate_const1}
\end{equation}
for the forward reaction under polariton condensation, where 
\begin{equation}
\begin{aligned}
k_{\text{R}\rightarrow \text{P}}^{\text{cond}}(n)&=\sqrt{\frac{\pi}{\lambda_{\text{S}}k_{\text{B}}T}}\frac{|J_{\text{R}\text{P}}|^{2}}{\hbar}\sum_{\nu_{+}=0}^{\infty}\sum_{\nu_{-}=0}^{\infty}\sum_{\nu_{\text{D}}=0}^{\infty}W_{\nu_{+},\nu_{-},\nu_{\text{D}}}^{f,n},\\
W_{\nu_{+},\nu_{-},\nu_{\text{D}}}^{f,n} & =|F_{\nu_{+},\nu_{-},\nu_{\text{D}}}^{f,n}|^{2}\times\exp(-\frac{E_{\nu_{+},\nu_{-},\nu_{\text{D}}}^{f,n\ddagger}}{k_{\text{B}}T}).\label{eq:rate_const2}
\end{aligned}
\end{equation}
The FC factor $|F_{\nu_{+},\nu_{-},\nu_{\text{D}}}^{f,n}|^{2}=|\bra{0,N_{-},n}\ket{\nu_{+},\nu_{-},\nu_{\text{D}}}'|^{2}$,
and activation energy $E_{\nu_{+},\nu_{-},\nu_{\text{D}}}^{f,n\ddagger}$ play
an important role in determining the rate constant.

While many methods have been developed for computing multimode FC
factors \cite{roche1990polyatomic,sharp1964franck,toniolo2001efficient}, the focus has been on increasing the number of modes while keeping
their occupation small. The current problem, however, offers a new
technical challenge: the large occupation of $\text{LP}$ makes the aforementioned
methods computationally expensive. Instead, we draw inspiration from
previous work that employs generating functions \cite{sharp1964franck} and combine those techniques with the powerful Lagrange-B\"urmann formula
\cite{whittaker1996course} to obtain analytical expressions
for the required three-dimensional FC factors (see details in Supplementary
Section S3). 

The activation energies for the various channels of reactivity
are, 
\begin{equation}
\begin{aligned}E_{\nu_{+},\nu_{-},\nu_{\text{D}}}^{f,n\ddagger} & =\frac{(E_{\text{P}}^{\nu_{+},\nu_{-},\nu_{\text{D}}}-E_{\text{R}}^{0,N_{-},n}+\lambda_{\text{S}})^{2}}{4\lambda_{\text{S}}},\end{aligned}
\end{equation}
where 
\begin{equation}
\begin{aligned}E_{\text{P}}^{\nu_{+},\nu_{-},\nu_{\text{D}}}= & \Delta G+\hbar\bigg[\omega_{+}'\Big(\nu_{+}+\frac{1}{2}\Big)\\
 & \hspace{30pt}+\omega_{-}'\Big(\nu_{-}+\frac{1}{2}\Big)+\omega_{\text{vib}}\Big(\nu_{\text{D}}+\frac{1}{2}\Big)\bigg],\\
E_{\text{R}}^{0,N_{-},n}= & \hbar\bigg[\omega_{+}\frac{1}{2}+\omega_{-}\Big(N_{-}+\frac{1}{2}\Big)+\omega_{\text{vib}}\Big(n+\frac{1}{2}\Big)\bigg].
\end{aligned}
\end{equation}

When condensation takes place, the number of quanta in the lower polariton
$N_{-}\sim10^{5}$ is so large that the summation in $k_{\text{R}\rightarrow \text{P}}^{\text{cond}}(n)$ becomes difficult to estimate. To simplify the computation and gain
intuition, we group channels into sets with same change in total number
of intramolecular vibrational quanta $f=\nu_{+}+\nu_{-}+\nu_{\text{D}}-N_{-}-n$
upon ET. The closeness in energy between PES with same
$f$, and hence similar activation barriers, is the rationale for this grouping. $k_{\text{R}\rightarrow \text{P}}^{\text{cond}}(n)$ then goes from a
free summation over three indices $\nu_{+}$, $\nu_{-}$ and $\nu_{\text{D}}$
into a summation over four indices $f$, $\nu_{+}$, $\nu_{-}$
and $\nu_{\text{D}}$ with the constraint $\nu_{+}+\nu_{-}+\nu_{\text{D}}=N_{-}+n+f$,
\begin{equation}
\begin{aligned}
k_{\text{R}\rightarrow \text{P}}^{\text{cond}}(n)=&\sqrt{\frac{\pi}{\lambda_{\text{S}}k_{\text{B}}T}}\frac{|J_{\text{R}\text{P}}|^{2}}{\hbar}\\
&\sum_{f=-N_{-}-n}^{\infty}\sum_{\nu_{+},\nu_{-},\nu_{\text{D}}}^{\nu_{+}+\nu_{-}+\nu_{\text{D}}=N_{-}+n+f}W_{\nu_{+},\nu_{-},\nu_{\text{D}}}^{f,n}.
\end{aligned}
\end{equation}
\begin{figure}[!h]
	\subfloat{ \includegraphics[width=0.45\textwidth]{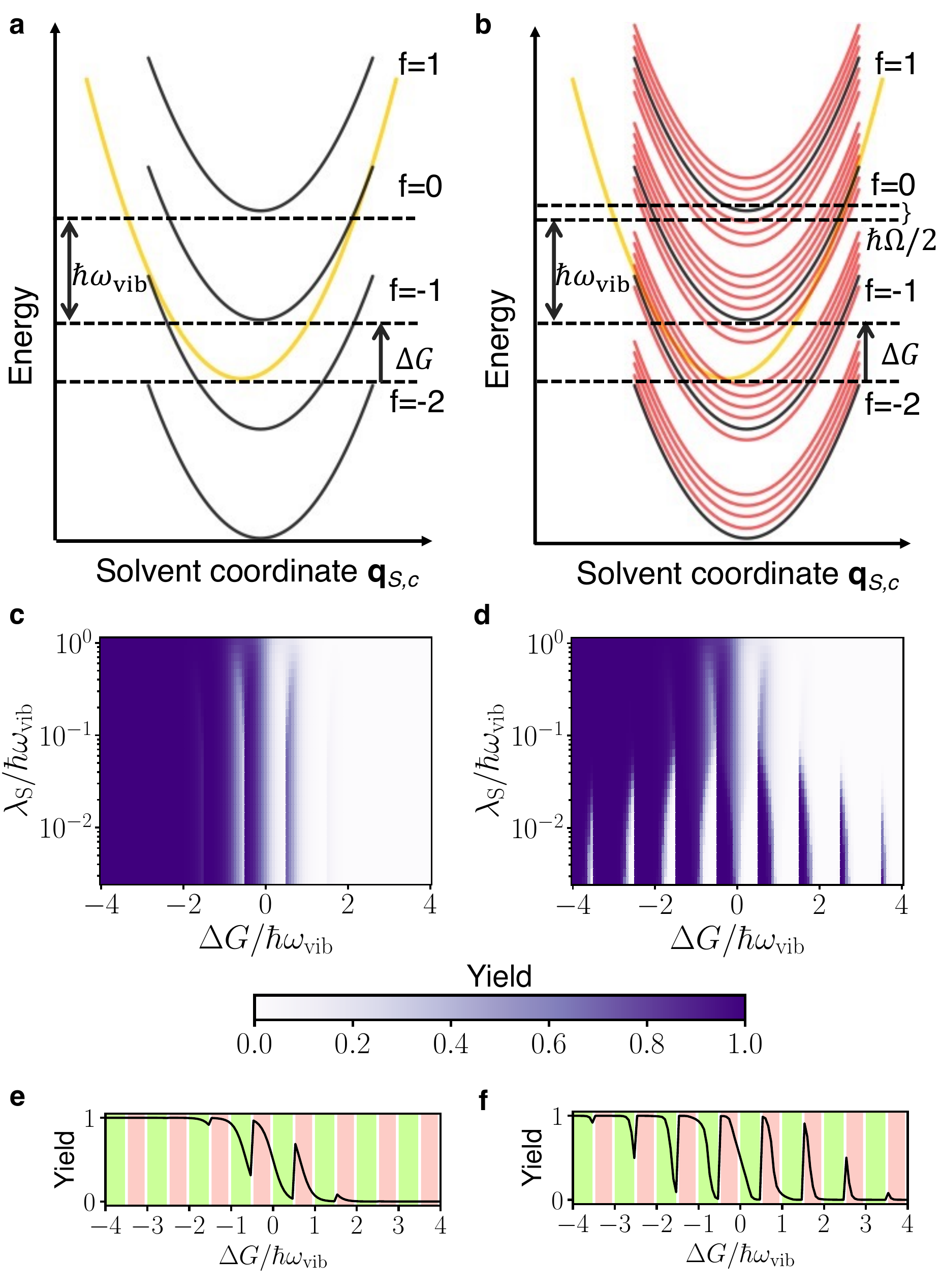}}\caption{\label{fig:yield} Potential energy surfaces (not to scale)
			and reaction yield. a, c, e are results for a laser driven system without light-matter strong coupling (SC) and b, d, f are for the same system under SC and $2.4\%$ of the population in the lower polariton (condensation). All these plots are for symmetric light-matter coupling $g_{\text{R}}=g_{\text{P}}$. a, b For a clearer qualitative picture, we plot the PESs under zero detuning $\Delta=0$. Initial (yellow) and final (black) PESs for a molecule undergoing the forward reaction with solvent coordinate $\mathbf{q_{\text{S},c}}$. While the energy separation between black PESs is $\hbar\omega_{\text{vib}}$, the condensate provides many additional final PESs (red, separated by $\hbar\Omega/2$ at resonance). c Reaction yield $N_{\text{P}}^{\text{ss}}/N$ at temperature $k_{\text{B}}T=0.1389\hbar\omega_{\text{vib}}$ ($T=298$K when $\hbar \omega_{\text{vib}}=185$ meV), Huang-Rhys factor $S=3.5$, and average occupation of the intramolecular vibrational mode $\bar{n}=0.08$. d Reaction yield $N_{\text{P}}^{\text{ss}}/N$ with $\Delta=-0.0695\omega_{\text{vib}}$, $2g_{\text{R}}\sqrt{N}=2g_{\text{P}}\sqrt{N}=0.1\omega_{\text{vib}}$, $P_{-}=0.08N\Gamma_{\downarrow}$, $N=10^7$, temperature and Huang-Rhys factor are the same as c. The contributions of the red PESs through the condensate provide a broader tunability of reaction yields with respect to $\Delta G$ than under laser driving without SC. Notice that originally endergonic (exergonic) reactions in the absence of optical pumping can become exergonic (endergonic) under the featured nonequilibrium conditions. e, f A cross-section of plot (c-d) when $\lambda_{\text{S}}=10^{-2}\hbar\omega_{\text{vib}}$. The pink shaded regions correspond to cases where the dominant forward (backward) channel is in the inverted (normal) regime; the opposite is true for the green shaded regions. The condensate amplifies the forward (backward) reaction in the pink (green) shaded regions.}
\end{figure}

To understand the qualitative difference between reactions under polariton condensation and external pumping without SC, in Fig  \ref{fig:yield}a-b we plot the PESs (not to scale) showing the forward reaction under symmetric light-matter coupling and zero detuning. The yellow (black) parabolas in Fig. \ref{fig:yield}a-b represent PESs for a molecule
in electronic state $\ket{\text{R}}$ ($\ket{\text{P}}$) and vibrational state $\ket{2}$ ($\ket{2+f}'$) in Fig.
\ref{fig:yield}a and $\ket{0,N_{-},2}$ ($\ket{0,N_{-},2+f}'$) in Fig. \ref{fig:yield}b. The red parabolas in Fig. \ref{fig:yield}b are additional final PESs provided by the vibrational polariton condensate (hereafter referred to solely as ``condensate") that account for all other final vibrational states $\ket{\nu_{+},\nu_{-},\nu_{\text{D}}}'$.

\begin{figure}
	\subfloat{ \includegraphics[width=0.45\textwidth]{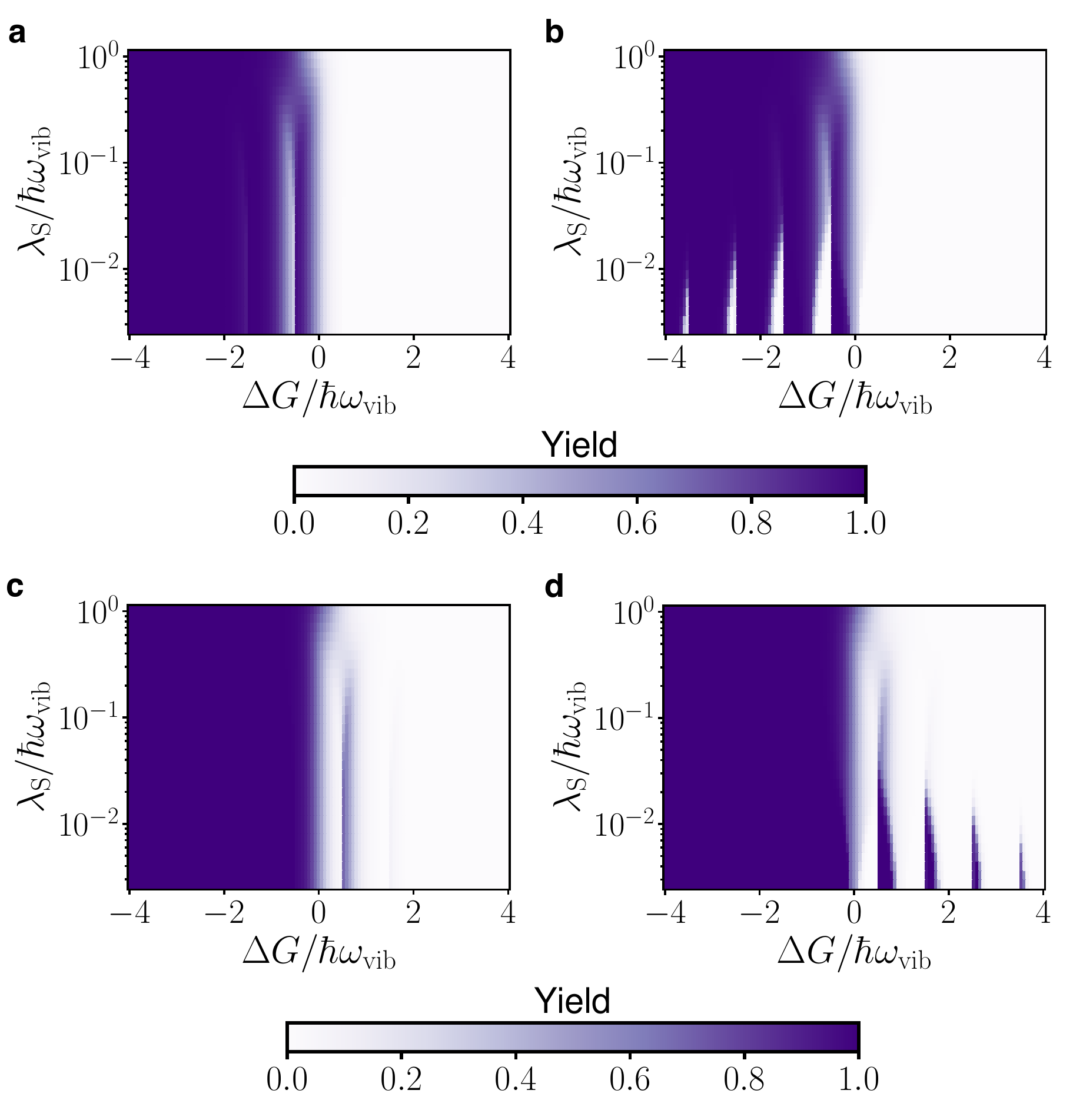}}\caption{\label{fig:Yieldrp}Reaction yield for asymmetric light-matter
			coupling. a (c) The yield of the reaction when only the product (reactant) weakly couples with light. b (d) Analogous plots under strong coupling $2g_{\text{P}}\sqrt{N}=0.1\omega_{\text{vib}}$, $g_{\text{R}}=0$ ($g_{\text{P}}=0$, $2g_{\text{R}}\sqrt{N}=0.1\omega_{\text{vib}}$). We use parameters $\Delta=-0.0695\omega_{\text{vib}}$, $k_{\text{B}}T=0.1389\hbar\omega_{\text{vib}}$ ($T=298$K when $\hbar \omega_{\text{vib}}=185$ meV), $S=3.5$, $P_{-}=0.08N\Gamma_{\downarrow}$ and $N=10^7$. We assume the same scattering parameters $W_{ij}$ and decay rates $\Gamma_{\downarrow},$ $\kappa$ as in Fig. \ref{fig:Pconst}.}
\end{figure}

The net rate of ET is, 
\begin{equation}
\begin{aligned}\frac{dN_{\text{R}}}{dt} & =-k_{\text{R}\rightarrow \text{P}}^{z}N_{\text{R}}+k_{\text{P}\rightarrow \text{R}}^{z}N_{\text{P}},\end{aligned}
\label{eq:modYield}
\end{equation}
where $k_{\text{R}\rightarrow \text{P}}^{z}$ and $k_{\text{P}\rightarrow R}^{z}$ ($z=\text{IR},\text{cond}$)
are the rate constants for the forward and backward reactions, respectively,
which are themselves functions of $N_{\text{R}}$ and $N_{\text{P}}$ when $g_{\text{R}}\neq g_{\text{P}}$.
We find the steady state solution $N_{\text{R}}^{\text{ss}}$ from this equation
and compute the reaction yield $N_{\text{P}}^{\text{ss}}/N$.

The difference in yield between the condensate and bare case is particularly large when $\lambda_{\text{S}}\ll\hbar\omega_{\text{vib}}<|\Delta G|$ (see Fig. \ref{fig:yield}c-d for symmetric coupling $g_{\text{R}}=g_{\text{P}}$).
To understand the underlying reason, we define the dominant channel
$f_{\text{min}}$ as the one with minimum activation barrier outside of the
cavity.
\begin{equation}
\frac{1}{k_{\text{B}}T}\frac{dE_{f}^{\ddagger}}{df}=\frac{\hbar\omega_{\text{vib}}}{k_{\text{B}}T}\bigg(\frac{\Delta G+\lambda_{\text{S}}+f\hbar\omega_{\text{vib}}}{2\lambda_{\text{S}}}\bigg)\label{eq:extremum}
\end{equation}
Setting the derivative in Eq. (\ref{eq:extremum}) equal to zero and
taking into account the discrete nature of $f$, we find the dominant
channel, $f_{\text{min}}=\Big\lceil\frac{-\Delta G-\lambda_{\text{S}}}{\hbar\omega_{\text{vib}}}\Big\rceil$
or $\Big\lfloor\frac{-\Delta G-\lambda_{\text{S}}}{\hbar\omega_{\text{vib}}}\Big\rfloor$.
When $\lambda_{\text{S}}\ll\hbar\omega_{\text{vib}},|\Delta G|$, this channel contributes
most to the rate constant because $\frac{1}{k_{\text{B}}T}\Big|\frac{dE_{f}^{\ddagger}}{df}\Big|\gg1$.
We define Marcus normal $\frac{dE_{f}^{\ddagger}}{df}\Big|_{f_{\text{min}}}>0$
and inverted $\frac{dE_{f}^{\ddagger}}{df}\Big|_{f_{\text{min}}}<0$ regimes
with respect to the dominant channel. If the dominant forward channel
is in the inverted regime, the dominant backward channel (which can
be found by replacing $\Delta G\rightarrow-\Delta G$ in Eq. (\ref{eq:extremum})) will be in the normal regime when $\lambda_{\text{S}}\ll\hbar\omega_{\text{vib}},|\Delta G|$.

\begin{figure}
	\subfloat{ \includegraphics[width=0.4\textwidth]{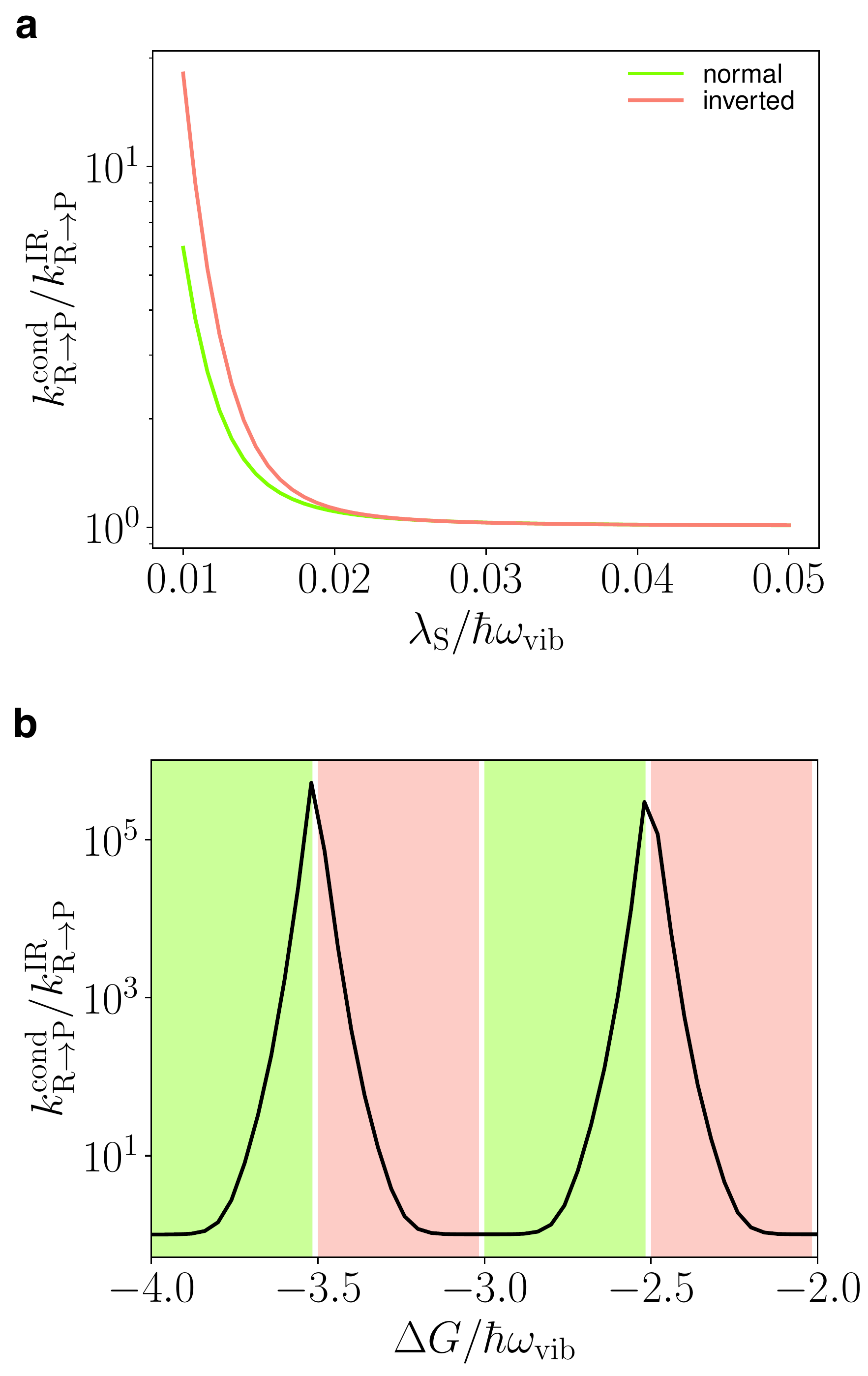}}\caption{\label{fig:rateconst}Rate constant. Ratio of the rate constants inside $k_{\text{R}\rightarrow \text{P}}^{\text{cond}}$ and outside $k_{\text{R}\rightarrow \text{P}}^{\text{IR}}$ of the cavity under laser excitation with $\Delta=-0.0695\omega_{\text{vib}}$, $k_{\text{B}}T=0.1389\hbar\omega_{\text{vib}}$ ($T=298$K when $\hbar \omega_{\text{vib}}=185$ meV), $S=3.5$, $2g\sqrt{N}=0.1\omega_{\text{vib}}$, $P_{-}=0.08N\Gamma_{\downarrow}$ and $N=10^7$ for symmetric coupling $g_{\text{R}}=g_{\text{P}}=g$. a Relative rate constant as a function of reorganization energy, $\lambda_{\text{S}}$, with $\Delta G= -3.33\hbar\omega_{\text{vib}}$ (the pink curve; here, the dominant channel lies in the inverted regime) and $\Delta G=-3.73\hbar\omega_{\text{vib}}$ (the green curve; here, the dominant channel lies in the normal regime) and b as a function of $\Delta G$ with $\lambda_{\text{S}}=10^{-2}\hbar\omega_{\text{vib}}$. Here, the pink region corresponds to the dominant channel in the inverted regime and the green to the normal regime.}
\end{figure}

In Fig. \ref{fig:yield}d, we see periodic yield modification in $\Delta G$ with period $\sim\hbar\omega_{\text{vib}}$, which decays for large $\Delta G/\hbar\omega_{\text{vib}}$ due to concomitant decline in FC factor for large changes in the number of vibrational quanta between the initial and final states. Outside of the cavity, we only see the first fringe (Fig. \ref{fig:yield}c). To observe the full periodic structure in yield, we need large occupation of higher vibrational states which requires very large pumping rates outside of the cavity. However, under polariton condensation, the macroscopic population of the lower polariton enables these interesting periodic features to be observed at room temperature with experimentally attainable pumping rates. Additionally, polariton condensation not only modifies the reaction yield under symmetric light-matter coupling strengths, as seen in Fig. \ref{fig:yield}, it also changes the yield when the reactant and product asymmetrically couple with light (Fig. \ref{fig:Yieldrp}).

Condensation provides many additional channels for the forward and
backward reactions (separated by $\sim\hbar\Omega/2$, see red curves
in Fig. \ref{fig:yield}b, showing only the forward channels at resonance $\Delta=0$) due
to transfer of quanta between the polariton and dark modes during the reaction. Condensation speeds up the reaction when the dominant channel is either in the inverted or in the normal regime (Fig. \ref{fig:rateconst}b). This is the case because there are additional channels with energy both higher (benefiting the inverted regime) and lower (benefiting the normal regime) than the dominant channel (see red curves in Fig. \ref{fig:yield}b). Apart from reduced activation energy, the additional channels provided by the condensate also have large enough FC factors to affect the rate
constant. Reaction channels that involve changes in the number of quanta in the $\text{LP}$ during the reaction have significantly larger FC factors ($\sim 10^{20}$ times) under condensation $N_{-}=0.1N$ than without any pumping $N_{-}=0$ (Fig. S1 in the supplementary information compares them). Changes in the rate constant as a function of $\lambda_{\text{S}}$ (Fig. \ref{fig:rateconst}a) and $\Delta G$ (Fig. \ref{fig:rateconst}b) are large for small $\lambda_{\text{S}}/\hbar\omega_{\text{vib}}$ and when $\Delta G/\hbar \omega_{\text{vib}}=n/2$ where $n$ is an odd integer since activation energy effects are large for these set of parameters.

\section*{Discussion}
Our result is a first step towards understanding the effect of Bose-Einstein
condensation of polaritons on chemical reactivity. We demonstrate
this effect using a simple electron transfer model (MLJ) with molecular
vibrations strongly coupled to light. In particular, we show that
one can counteract the massive degeneracy of dark modes and enhance
polaritonic effects by having a macroscopic occupation of the lower
polariton mode \textit{i.e.}, Bose-Einstein condensation. Our results indicate that the latter is feasible for experimentally realizable pump powers
and Rabi splittings, despite the close proximity in energy of the
dark state manifold with $\hbar\Omega\sim k_{\text{B}}T$. These results
can guide the choice of suitable materials for condensation under
VSC. While laser driving without SC modifies the reaction yield, this
change is amplified by the condensate, due to the availability of
many additional reactive channels that differ in energy by $\sim\hbar\Omega/2$
rather than $\sim\hbar\omega_{\text{vib}}$. For a wide range of parameters,
we find that this leads to a periodic dependence of reaction yield
as a function of $\Delta G$ (with period $\sim\hbar\omega_{\text{vib}}$),
rendering a set of originally endergonic reactions exergonic, and
vice versa. These effects are substantially weaker under laser driving,
and highlight both the energetic (availability of additional channels
with lower activation energy) and entropic (redistribution of vibrational
energy from the condensate into the polariton and dark modes upon
reaction) advantages of exploiting polariton condensates for reactivity. To summarize, vibrational polariton condensation offers a novel strategy to accumulate energy into a well defined normal mode, a holy-grail in the field of vibrational dynamics that has been historically hindered by IVR. Its successful demonstration could revive hopes of ``mode selective chemistry" \cite{frei1985infrared}, beyond electron transfer processes. In future work, it
will be interesting to explore how the studied phenomena generalize to molecular polariton condensates in different spectral ranges.

\section*{Data availability}
Datasets generated by our code are available by email upon request to the authors.

\section*{Code availability}
Computational scripts used to generate the plots in the present article are available by email upon request to the authors.

\section*{acknowledgements}
S.P.S., L.A.M.M., and J.Y.Z. were supported by the US Department of Energy, Office of Science, Basic Energy Sciences, CPIMS Program under Early Career Research Program Award DE-SC0019188. J.A.C.G.A. was supported through AFOSR award FA9550-18-1-0289. This work used the Extreme Science and Engineering Discovery Environment (XSEDE), which is supported by National Science Foundation grant number ACI-1548562, under allocation number TGASC150024. J.Y.Z. acknowledges fruitful discussions with Wei Xiong. S.P.S thanks Juan P\'{e}rez-S\'{a}nchez and Matthew Du for helpful discussions.

\section*{Author contributions}
S.P.S. developed model, calculations, and interpretation of the results in the manuscript. L.A.M.M. provided guidance in the development of the initial model. J.A.C.G.A. developed the optimal basis to carry out the calculations for the electron transfer reaction under condensation and provided guidance on the interpretation of phenomenology. S.S. assisted on the calculation of multidimensional Franck-Condon factors. J.Y.Z. conceived the original version of the project and supervised the work throughout.

\section*{Competing interests}
The authors declare no competing interests.

\end{document}


\title{Supplementary Information: Driving chemical reactions with polariton condensates}

\author{Sindhana Pannir-Sivajothi}
\affiliation{Department of Chemistry and Biochemistry, University of California San Diego, La Jolla, California 92093, USA}
\author{Jorge A. Campos-Gonzalez-Angulo}
\affiliation{Department of Chemistry and Biochemistry, University of California San Diego, La Jolla, California 92093, USA}
\author{Luis A. Mart\'{i}nez-Mart\'{i}nez}
\affiliation{Department of Chemistry and Biochemistry, University of California San Diego, La Jolla, California 92093, USA}
\author{Shubham Sinha}
\affiliation{Department of Mathematics, University of California San Diego, La Jolla, California 92093, USA}
\author{Joel Yuen-Zhou}
\email{joelyuen@ucsd.edu}
\affiliation{Department of Chemistry and Biochemistry, University of California San Diego, La Jolla, California 92093, USA}
	\maketitle
	\subsection*{Supplementary Note 1}
	We use a generalization of the dark state basis introduced in \cite{strashko2016raman,campos2019resonant} to reduce the number of vibrational modes involved in the reaction. The bosonic operator for $\text{D}_{\text{R},c}$, the dark mode highly localized in the $c^{th}$ molecule when it is in electronic state $\ket{R}$ with total number of reactants $N_{\text{R}}$ and products $N_{\text{P}}$ is
	\begin{equation}
	\begin{aligned}
	\hat{a}_{\text{D}}^{(\text{R},c)}=&\sqrt{\frac{g_{\text{P}}^2N_{\text{P}}+g_{\text{R}}^2(N_{\text{R}}-1)}{g_{\text{P}}^2N_{\text{P}}+g_{\text{R}}^2N_{\text{R}}}}\hat{a}_{\text{R},c}-\frac{g_{\text{R}}^2}{\sqrt{g_{\text{P}}^2N_{\text{P}}+g_{\text{R}}^2N_{\text{R}}}}\frac{1}{\sqrt{g_{\text{P}}^2N_{\text{P}}+g_{\text{R}}^2(N_{\text{R}}-1)}}\sum_{i\neq c}^{N_{\text{R}}}\hat{a}_{\text{R},i}\\
	&-\frac{g_{\text{R}}g_{\text{P}}}{\sqrt{g_{\text{P}}^2N_{\text{P}}+g_{\text{R}}^2N_{\text{R}}}}\frac{1}{\sqrt{g_{\text{P}}^2N_{\text{P}}+g_{\text{R}}^2(N_{\text{R}}-1)}}\sum_{j=1}^{N_{\text{P}}}\hat{a}_{\text{P},j}.
	\end{aligned}
	\end{equation}
	Notice that, when $g_{\text{P}}=0$, this dark mode will not involve vibrations in the product molecules and when $g_{\text{R}}=0$, this dark mode will be the same as a vibration localized in the $c^{th}$ molecule $\hat{a}_{\text{D}}^{(\text{R},c)}=\hat{a}_{\text{R},c}$. Similarly, the bosonic operator for $\text{D}_{\text{P},c}'$, the dark mode highly localized in the $c^{th}$ molecule after it has reacted and is in electronic state $\ket{\text{P}}$ with total number of reactants $N_{\text{R}}-1$ and products $N_{\text{P}}+1$ is
	\begin{equation}
	\label{eq:dP}
	\begin{aligned}
	\hat{a}_{\text{D}}^{(\text{P},c)\prime}=&\sqrt{\frac{g_{\text{P}}^2N_{\text{P}}+g_{\text{R}}^2(N_{\text{R}}-1)}{g_{\text{P}}^2(N_{\text{P}}+1)+g_{\text{R}}^2(N_{\text{R}}-1)}}\hat{a}_{\text{P},c}-\frac{g_{\text{R}}g_{\text{P}}}{\sqrt{g_{\text{P}}^2(N_{\text{P}}+1)+g_{\text{R}}^2(N_{\text{R}}-1)}}\frac{1}{\sqrt{g_{\text{P}}^2N_{\text{P}}+g_{\text{R}}^2(N_{\text{R}}-1)}}\sum_{i=1}^{N_{\text{R}}-1}\hat{a}_{\text{R},i}\\
	&-\frac{g_{\text{P}}^2}{\sqrt{g_{\text{P}}^2(N_{\text{P}}+1)+g_{\text{R}}^2(N_{\text{R}}-1)}}\frac{1}{\sqrt{g_{\text{P}}^2N_{\text{P}}+g_{\text{R}}^2(N_{\text{R}}-1)}}\sum_{j\neq c}^{N_{\text{P}}+1}\hat{a}_{\text{P},j}.
	\end{aligned}
	\end{equation}
	Here, when $g_{\text{P}}=0$, the dark mode in equation (\ref{eq:dP}) will be the same as a vibration localized in the $c^{th}$ molecule $\hat{a}_{\text{D}}^{(\text{P},c)\prime}=\hat{a}_{\text{P},c}$ and when $g_{\text{R}}=0$, this dark mode will not involve vibrations in the reactant molecules.
	\subsection*{Supplementary Note 2}
	We use Boltzmann rate equations as in \cite{banyai2000condensation,cao2004condensation} to model polariton relaxation, and solve for the steady state of $N+1$ coupled differential equations. We assume that the scattering rate $W_{ij}$ between polariton and dark modes is the same for all dark modes, labeled by $k$, this gives $W_{\text{D}_{k}+}=W_{\text{D}+}$ and $W_{-\text{D}_k}=W_{-\text{D}}$ \cite{del2015quantum}. Since our interests lie in the distribution of energy between polariton and dark modes rather than individual dark modes, we can simplify the problem by summing over all dark-mode equations and considering only their total population $n_{\text{D}}=\sum_{k=2}^{N} n_{\text{D}}^k$. We have the following rate equations for populations in the lower $n_{-}$, upper $n_{+}$ polaritons and all dark modes $n_{\text{D}}$,
	\begin{equation}
	\begin{aligned}
	\frac{dn_{-}}{dt}&=R_{-\text{D}}+R_{-+}-\gamma_{-}n_{-}+P_{-},\\
	\frac{dn_{\text{D}}}{dt}&=-R_{-\text{D}} + R_{\text{D}+}-\gamma_{\text{D}}^kn_{\text{D}},\\
	\frac{dn_{+}}{dt}&=-R_{-+}-R_{\text{D}+}-\gamma_{+}n_{+},
	\end{aligned}
	\end{equation}
	where
	\begin{equation}
	\begin{aligned}
	R_{-\text{D}}=&W_{-\text{D}_k}\Big(n_{\text{D}}(1+n_{-})-e^{-\beta\hbar(\Omega-\Delta) /2}(N-1+n_{\text{D}})n_{-}\Big),\\
	R_{-+}=&W_{-+}\Big(n_{+}(1+n_{-})-e^{-\beta\hbar\Omega }(1+n_{+})n_{-}\Big),\\
	R_{\text{D}+}=&W_{\text{D}_k+}\Big(n_{+}(N-1+n_{\text{D}})-e^{-\beta\hbar(\Omega+\Delta)/2}(1+n_{+})n_{\text{D}}\Big).
	\end{aligned}
	\end{equation}
   
   The rate coefficients can be expressed as $W_{ij}=\alpha(1+n(\beta E_{ji}))$ and $W_{ji}=\alpha n(\beta E_{ji})$ where $\alpha$ is a temperature independent constant, $n(\beta E_{ji})$ is the average Bose-Einstein population at energy $E_{ji}=E_j-E_i$ where $E_j>E_i$ and inverse temperature $\beta=1/k_{\text{B}}T$. The rate coefficients $W_{ij}$ should also depend on the low-frequency vibrational density of states, for simplicity, we take the spectral density to be flat.

  For all calculations in the main manuscript, we use $\kappa=\Gamma_{\downarrow}$, $N=10^7$ and $\alpha=4.33\times 10^{-6}\Gamma_{\downarrow}$ which corresponds to $(N-1)W_{\text{D}_k-}=100\Gamma_{\downarrow}$ ($\sim$ 1 ps) at room temperature for light-matter coupling strength $2\hbar g\sqrt{N}=18.5\text{meV}$ at zero detuning $\Delta=0$. These values are similar to those in experiments $\kappa=10^{10}$ s$^{-1}$ ($\sim 100$ ps), $\Gamma_{\downarrow}= 10^{10}$ s$^{-1}$ ($\sim 100$ ps), and scattering from $\text{LP}$ to all dark modes $(N-1)W_{\text{D}_k-}= 10^{12}$ $s^{-1}$ ($\sim 1$ ps) \cite{xiang2019state}. 

\subsection*{Supplementary Note 3}
\begin{figure}[!h]
	\subfloat{\includegraphics[width=0.45\textwidth]{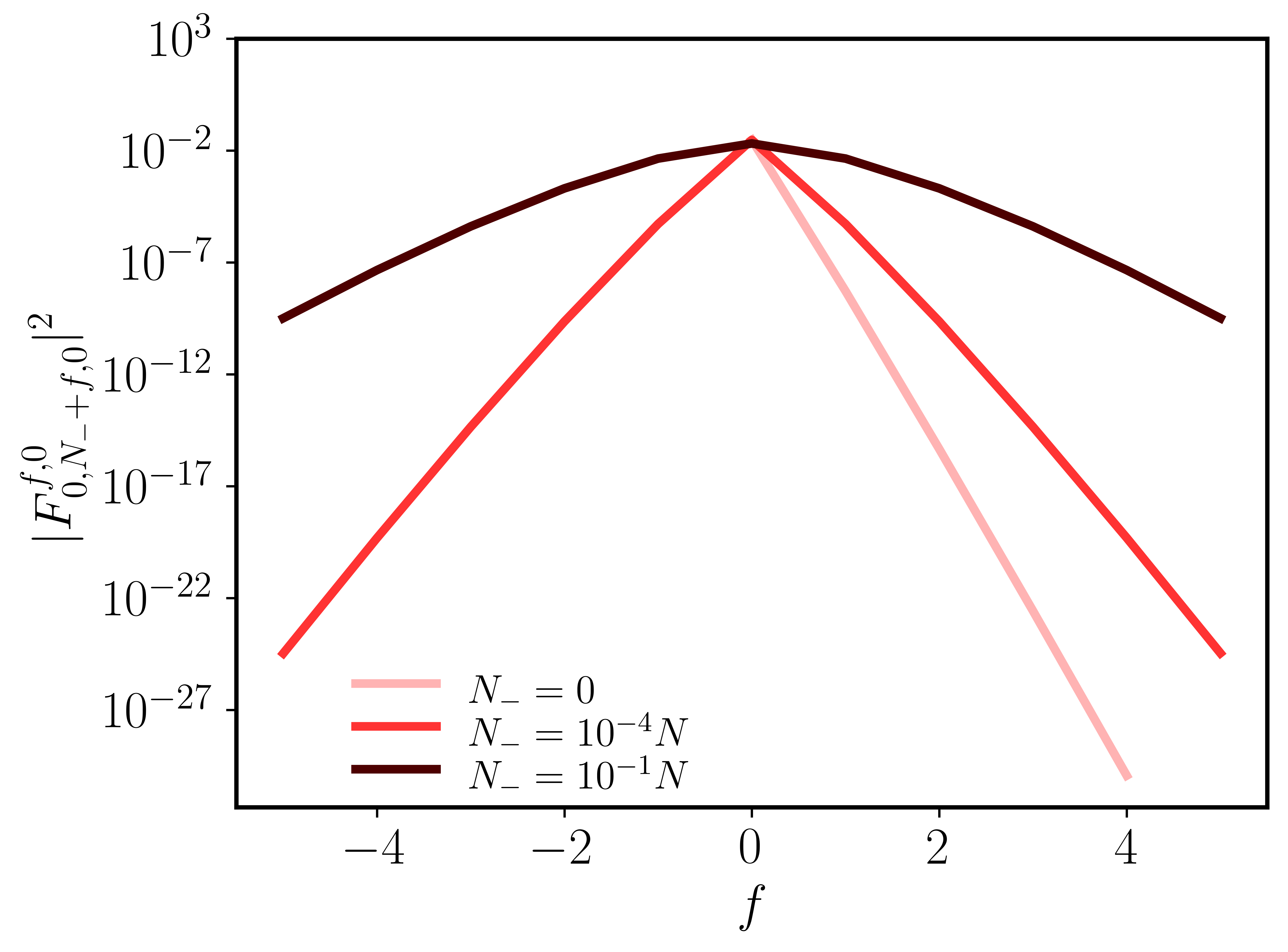}}\caption{\label{fig:FC}\textbf{Franck-Condon (FC) factors for different channels
			as a function of $\text{LP}$ population.} Reaction rate changes by the condensate
		occur due to channels featuring gain/loss $f$ in vibrational quanta
		in the $\text{LP}$. The contribution of these channels is proportional to the
		corresponding FC factors and becomes more significant as the $\text{LP}$ population
		$N_{-}$ increases. Here, we show FC factors for $N_{-}=0,10^{-4}N,10^{-1}N$
		at resonance $\Delta=0$ with symmetric light-matter coupling $2g_{\text{R}}\sqrt{N}=2g_{\text{P}}\sqrt{N}=0.1\omega_{vib}$ and Huang-Rhys factor $S=3.5$.}
\end{figure}

Dependence of the FC factors on $\text{LP}$ population is shown
in Fig. \ref{fig:FC} and they scale much better when the population in $\text{LP}$ is large. We obtain an analytical expression for the Franck-Condon factor $|F_{\nu_+,\nu_-,\nu_{\text{D}}}^{f,0}|^2$,
\begin{equation}
\begin{aligned}
F_{\nu_+,\nu_-,\nu_{\text{D}}}^{f,0}=&\bra{0,N_{-},0}\ket{\nu_+,\nu_-,\nu_{\text{D}}}'\\
=&
\begin{cases}
\sqrt{e^{-S}S^f}\sqrt{\frac{N_{-}!}{\nu_+!\nu_-!\nu_{\text{D}}!}}\tilde{w}^{-f}\Big[q^{N_{-}}\Big]\Bigg((\tilde{x}+qx)^{\nu_+}(\tilde{y}+yq)^{\nu_-}(\tilde{z}-\tilde{z}q+zq)^{\nu_{\text{D}}}\frac{\exp(\frac{-Swq}{1-q-wq})}{(1-q-wq)^{f+1}}\Bigg)\hspace{20pt}&g_{\text{P}}\neq 0\\
(-1)^{\nu_{\text{D}}}\sqrt{\frac{N_{-}!}{\nu_+!\nu_-!(\nu_{\text{D}}-f)!}}(x+\tilde{x})^{\nu_+}(y+\tilde{y})^{\nu_-}u^{\nu_{\text{D}}-f}\bra{(\nu_{\text{D}}-f)_{\text{R},c}}\ket{(\nu_{\text{D}})_{\text{P},c}}\hspace{20pt}&g_{\text{P}}=0
\end{cases}
\end{aligned}\label{eq:fcfactors}
\end{equation}
where $[q^{n}]G(q)$ is the coefficient of $q^n$ when you expand $G(q)$ as a Taylor series, and
\begin{equation}
\begin{aligned}
u&=\Bigg(-\cos\theta\frac{g_{\text{R}}}{\sqrt{g_{\text{R}}^2N_{\text{R}}+g_{\text{P}}^2N_{\text{P}}}}\Bigg),\\
x&=\sin\theta\cos\theta',\\
y&=\sin\theta\sin\theta',\\
z&=\Bigg(g_{\text{P}}\cos\theta\sqrt{\frac{g_{\text{R}}^2(N_{\text{R}}-1)+g_{\text{P}}^2N_{\text{P}}}{(g_{\text{R}}^2(N_{\text{R}}-1)+g_{\text{P}}^2(N_{\text{P}}+1))(g_{\text{R}}^2N_{\text{R}}+g_{\text{P}}^2N_{\text{P}})}}\Bigg),\\
w&=\Bigg(\frac{g_{\text{R}}g_{\text{P}}}{g_{\text{R}}^2(N_{\text{R}}-1)+g_{\text{P}}^2N_{\text{P}}}\Bigg),\\
\tilde{x}&=\Bigg(\frac{-\cos\theta\sin\theta'(g_{\text{R}}^2(N_{\text{R}}-1)+g_{\text{P}}^2N_{\text{P}})}{\sqrt{g_{\text{R}}^2N_{\text{R}}+g_{\text{P}}^2N_{\text{P}}}\sqrt{g_{\text{R}}^2(N_{\text{R}}-1)+g_{\text{P}}^2(N_{\text{P}}+1)}}\Bigg),\\
\tilde{y}&=\Bigg(\frac{\cos\theta \cos\theta'(g_{\text{R}}^2(N_{\text{R}}-1)+g_{\text{P}}^2N_{\text{P}})}{\sqrt{g_{\text{R}}^2N_{\text{R}}+g_{\text{P}}^2N_{\text{P}}}\sqrt{g_{\text{R}}^2(N_{\text{R}}-1)+g_{\text{P}}^2(N_{\text{P}}+1)}}\Bigg),\\
\tilde{z}&=\Bigg(-\cos\theta\frac{g_{\text{R}}^2(N_{\text{R}}-1)+g_{\text{P}}^2N_{\text{P}}}{g_{\text{P}}\sqrt{g_{\text{R}}^2N_{\text{R}}+g_{\text{P}}^2N_{\text{P}}}}\sqrt{\frac{g_{\text{R}}^2(N_{\text{R}}-1)+g_{\text{P}}^2N_{\text{P}}}{g_{\text{R}}^2(N_{\text{R}}-1)+g_{\text{P}}^2(N_{\text{P}}+1)}}\Bigg),\\
\tilde{w}&=\Bigg(-\cos\theta\frac{g_{\text{R}}^2(N_{\text{R}}-1)+g_{\text{P}}^2N_{\text{P}}}{g_{\text{P}}\sqrt{g_{\text{R}}^2N_{\text{R}}+g_{\text{P}}^2N_{\text{P}}}}\Bigg).
\end{aligned}\label{eq:xyz}
\end{equation}
We arrived at the expression in equation (\ref{eq:fcfactors}) by using generating functions and the Lagrange-B\"urmann formula \cite{whittaker1996course}. We then recursively compute $F_{\nu_+,\nu_-,\nu_{\text{D}}}^{f,n}$ from $F_{\nu_+,\nu_-,\nu_{\text{D}}}^{f,0}$ \cite{roche1990polyatomic}. Using
\begin{equation}
\begin{aligned}
\begin{bmatrix}
\hat{a}_+ \\    \hat{a}_- \\    \hat{a}_{\text{D}}^{(\text{R},c)}
\end{bmatrix}=
\begin{bmatrix}
J_{11} & J_{12} & J_{13}\\
J_{21} & J_{22} & J_{23}\\
J_{31} & J_{32} & J_{33}
\end{bmatrix}
\begin{bmatrix}
\hat{a}_+'\\
\hat{a}_-'\\
\hat{a}_{\text{D}}^{(\text{P},c)\prime}
\end{bmatrix}
+\begin{bmatrix}
K_1\\
K_2\\
K_3
\end{bmatrix}
\end{aligned}
\end{equation}
where
\begin{equation}
\begin{aligned}
J_{11}=&\cos\theta\cos\theta'+\sin\theta\sin\theta'\Bigg(\frac{g_{\text{R}}^2(N_{\text{R}}-1)+g_{\text{P}}^2N_{\text{P}}+g_{\text{R}}g_{\text{P}}}{\sqrt{g_{\text{R}}^2N_{\text{R}}+g_{\text{P}}^2N_{\text{P}}}\sqrt{g_{\text{R}}^2(N_{\text{R}}-1)+g_{\text{P}}^2(N_{\text{P}}+1)}}\Bigg)\\
J_{12}=&\sin\theta'\cos\theta-\cos\theta'\sin\theta\Bigg(\frac{g_{\text{R}}^2(N_{\text{R}}-1)+g_{\text{P}}^2N_{\text{P}}+g_{\text{R}}g_{\text{P}}}{\sqrt{g_{\text{R}}^2N_{\text{R}}+g_{\text{P}}^2N_{\text{P}}}\sqrt{g_{\text{R}}^2(N_{\text{R}}-1)+g_{\text{P}}^2(N_{\text{P}}+1)}}\Bigg)\\
J_{13}=&\sin\theta\Bigg(\frac{(g_{\text{R}}-g_{\text{P}})\sqrt{g_{\text{R}}^2(N_{\text{R}}-1)+g_{\text{P}}^2N_{\text{P}}}}{\sqrt{g_{\text{R}}^2N_{\text{R}}+g_{\text{P}}^2N_{\text{P}}}\sqrt{g_{\text{R}}^2(N_{\text{R}}-1)+g_{\text{P}}^2(N_{\text{P}}+1)}}\Bigg)\\
J_{21}=&\cos\theta'\sin\theta-\sin\theta'\cos\theta\Bigg(\frac{g_{\text{R}}^2(N_{\text{R}}-1)+g_{\text{P}}^2N_{\text{P}}+g_{\text{R}}g_{\text{P}}}{\sqrt{g_{\text{R}}^2N_{\text{R}}+g_{\text{P}}^2N_{\text{P}}}\sqrt{g_{\text{R}}^2(N_{\text{R}}-1)+g_{\text{P}}^2(N_{\text{P}}+1)}}\Bigg)\\
J_{22}=&\sin\theta\sin\theta'+\cos\theta\cos\theta'\Bigg(\frac{g_{\text{R}}^2(N_{\text{R}}-1)+g_{\text{P}}^2N_{\text{P}}+g_{\text{R}}g_{\text{P}}}{\sqrt{g_{\text{R}}^2N_{\text{R}}+g_{\text{P}}^2N_{\text{P}}}\sqrt{g_{\text{R}}^2(N_{\text{R}}-1)+g_{\text{P}}^2(N_{\text{P}}+1)}}\Bigg)\\
J_{23}=&-\cos\theta\Bigg(\frac{(g_{\text{R}}-g_{\text{P}})\sqrt{g_{\text{R}}^2(N_{\text{R}}-1)+g_{\text{P}}^2N_{\text{P}}}}{\sqrt{g_{\text{R}}^2N_{\text{R}}+g_{\text{P}}^2N_{\text{P}}}\sqrt{g_{\text{R}}^2(N_{\text{R}}-1)+g_{\text{P}}^2(N_{\text{P}}+1)}}\Bigg)
\end{aligned}
\end{equation}
and
\begin{equation}
\begin{aligned}
J_{31}=&-\sin\theta'\Bigg(\frac{(g_{\text{R}}-g_{\text{P}})\sqrt{g_{\text{R}}^2(N_{\text{R}}-1)+g_{\text{P}}^2N_{\text{P}}}}{\sqrt{g_{\text{R}}^2N_{\text{R}}+g_{\text{P}}^2N_{\text{P}}}\sqrt{g_{\text{R}}^2(N_{\text{R}}-1)+g_{\text{P}}^2(N_{\text{P}}+1)}}\Bigg)\\
J_{32}=&\cos\theta'\Bigg(\frac{(g_{\text{R}}-g_{\text{P}})\sqrt{g_{\text{R}}^2(N_{\text{R}}-1)+g_{\text{P}}^2N_{\text{P}}}}{\sqrt{g_{\text{R}}^2N_{\text{R}}+g_{\text{P}}^2N_{\text{P}}}\sqrt{g_{\text{R}}^2(N_{\text{R}}-1)+g_{\text{P}}^2(N_{\text{P}}+1)}}\Bigg)\\
J_{33}=&\Bigg(\frac{g_{\text{R}}^2(N_{\text{R}}-1)+g_{\text{P}}^2N_{\text{P}}+g_{\text{R}}g_{\text{P}}}{\sqrt{g_{\text{R}}^2N_{\text{R}}+g_{\text{P}}^2N_{\text{P}}}\sqrt{g_{\text{R}}^2(N_{\text{R}}-1)+g_{\text{P}}^2(N_{\text{P}}+1)}}\Bigg)\\
K_{1}=&-\sin\theta\frac{g_{\text{R}}}{\sqrt{g_{\text{R}}^2N_{\text{R}}+g_{\text{P}}^2N_{\text{P}}}}\sqrt{S}\\
K_{2}=&\cos\theta\frac{g_{\text{R}}}{\sqrt{g_{\text{R}}^2N_{\text{R}}+g_{\text{P}}^2N_{\text{P}}}}\sqrt{S}\\
K_{3}=&-\sqrt{\frac{g_{\text{R}}^2(N_{\text{R}}-1)+g_{\text{P}}^2N_{\text{P}}}{g_{\text{R}}^2N_{\text{R}}+g_{\text{P}}^2N_{\text{P}}}}\sqrt{S}
\end{aligned}
\end{equation}
we get the recursive formula
\begin{equation}
F_{\nu_+,\nu_-,\nu_{\text{D}}}^{f,n+1}=\frac{1}{\sqrt{n+1}}\Big(J_{31}\sqrt{\nu_+}F_{\nu_+-1,\nu_-,\nu_{\text{D}}}^{f,n}+F^{f,n}_{\nu_+,\nu_--1,\nu_{\text{D}}}J_{32}\sqrt{\nu_-}+J_{33}\sqrt{\nu_{\text{D}}}F_{\nu_+,\nu_-,\nu_{\text{D}}-1}^{f,n}+K_{3}F_{\nu_+,\nu_-,\nu_{\text{D}}}^{f-1,n}\Big).
\end{equation}

To derive equation (\ref{eq:fcfactors}), we write the initial and final vibrational states in terms of creation/annihilation operators for the upper, lower polaritons and dark modes
\begin{equation}
F_{\nu_+,\nu_-,\nu_{\text{D}}}^{f,0}
=\bra{0_{+}0_{-}0_{\text{D}}^{(\text{R},c)}}\frac{\Big(\hat{a}_{-}\Big)^{N_{-}}}{\sqrt{N_{-}!}}\frac{\Big(\hat{a}^{\prime\dagger}_{+}\Big)^{\nu_+}}{\sqrt{\nu_+!}}\frac{\Big(\hat{a}^{\prime\dagger}_{-}\Big)^{\nu_-}}{\sqrt{\nu_-!}}\frac{\Big(\hat{a}^{(\text{P},c)\prime\dagger}_{\text{D}}\Big)^{\nu_{\text{D}}}}{\sqrt{\nu_{\text{D}}!}}\ket{0_{+}'0_{-}'0^{(\text{P},c)\prime}_{\text{D}}}.
\end{equation}
Writing the polariton modes as a linear combination of the photon $\hat{a}_{\text{ph}}$ and bright modes $\hat{a}_{\text{B}(N_{\text{R}},N_{\text{P}})}$, $\hat{a}_{\text{B}(N_{\text{R}}-1,N_{\text{P}}+1)}$,
\begin{equation}
\begin{aligned}
F_{\nu_+,\nu_-,\nu_{\text{D}}}^{f,0}=&\frac{1}{\sqrt{N_{-}!\nu_+!\nu_-!\nu_{\text{D}}!}}\bra{0_{\text{ph}}0_{\text{B}(N_{\text{R}},N_{\text{P}})}0^{(\text{R},c)}_{\text{D}}}\Big(\sin\theta\hat{a}_{\text{ph}}-\cos\theta\hat{a}_{\text{B}(N_{\text{R}},N_{\text{P}})}\Big)^{N_{-}}\Big(\cos\theta'\hat{a}_{\text{ph}}^{\dagger}+\sin\theta'\hat{a}_{\text{B}(N_{\text{R}}-1,N_{\text{P}}+1)}^{\dagger}\Big)^{\nu_+}\\
&\times\Big(\sin\theta'\hat{a}_{\text{ph}}^{\dagger}-\cos\theta'\hat{a}_{\text{B}(N_{\text{R}}-1,N_{\text{P}}+1)}^{\dagger}\Big)^{\nu_-}\Big(\hat{a}^{(\text{P},c)\prime\dagger}_{\text{D}}\Big)^{\nu_{\text{D}}}\ket{0_{\text{ph}}0_{\text{B}(N_{\text{R}}-1,N_{\text{P}}+1)}0^{(\text{P},c)\prime}_{\text{D}}}.
\end{aligned}
\end{equation}
Here, $\comm{\hat{a}_{\text{ph}}}{\hat{a}_{\text{B}(N_{\text{R}},N_{\text{P}})}}=0$ and $\comm{\hat{a}_{\text{ph}}}{\hat{a}_{\text{B}(N_{\text{R}}-1,N_{\text{P}}+1)}}=0$, so we can use the binomial theorem and collect these operators,
\begin{equation}
\begin{aligned}
F_{\nu_+,\nu_-,\nu_{\text{D}}}^{f,0}=&\frac{1}{\sqrt{N_{-}!\nu_+!\nu_-!\nu_{\text{D}}!}}\sum_{l=0}^{N_{-}}\sum_{m=0}^{\nu_+}\sum_{n=0}^{\nu_-}\binom{N_{-}}{l}\binom{\nu_+}{m}\binom{\nu_-}{n}\bra{0_{\text{ph}}0_{\text{B}(N_{\text{R}},N_{\text{P}})}0^{(\text{R},c)}_{\text{D}}}\Big(\sin\theta\hat{a}_{\text{ph}}\Big)^l\\
&\times\Big(-\cos\theta\hat{a}_{\text{B}(N_{\text{R}},N_{\text{P}})}\Big)^{N_{-}-l}\Big(\cos\theta'\hat{a}_{\text{ph}}^{\dagger}\Big)^m\Big(\sin\theta'\hat{a}_{\text{B}(N_{\text{R}}-1,N_{\text{P}}+1)}^{\dagger}\Big)^{\nu_+-m}\Big(\sin\theta'\hat{a}_{\text{ph}}^{\dagger}\Big)^n\\
&\times\Big(-\cos\theta'\hat{a}_{\text{B}(N_{\text{R}}-1,N_{\text{P}}+1)}^{\dagger}\Big)^{\nu_--n}\Big(\hat{a}^{(\text{P},c)\prime\dagger}_{\text{D}}\Big)^{\nu_{\text{D}}}\ket{0_{\text{ph}}0_{\text{B}(N_{\text{R}}-1,N_{\text{P}}+1)}0^{(\text{P},c)\prime}_{\text{D}}}.
\end{aligned}
\end{equation}
The only non-vanishing terms in the above summation will be those with equal number of photon creation and annihilation operators $l=m+n$, since the photon mode does not get displaced during the chemical reaction, the overlap is non-zero only when the initial and final states are exactly the same. Plugging in $l=m+n$ and $\bra{0_{\text{ph}}}(\hat{a}_{\text{ph}})^{l}(\hat{a}_{\text{ph}}^{\dagger})^{m+n}\ket{0_{\text{ph}}}=(m+n)!$
\begin{equation}
\begin{aligned}
F_{\nu_+,\nu_-,\nu_{\text{D}}}^{f,0}=&\frac{1}{\sqrt{N_{-}!\nu_+!\nu_-!\nu_{\text{D}}!}}\sum_{m=0}^{\nu_+}\sum_{n=0}^{\nu_-}\binom{N_{-}}{m+n}\binom{\nu_+}{m}\binom{\nu_-}{n}\Big(\sin\theta\Big)^{m+n}\Big(\cos\theta'\Big)^m\Big(\sin\theta'\Big)^n\\
&\times\Big(-\cos\theta\Big)^{N_{-}-m-n}\Big(\sin\theta'\Big)^{\nu_+-m}\Big(-\cos\theta'\Big)^{\nu_--n}(m+n)!\\
&\times\bra{0_{\text{B}(N_{\text{R}},N_{\text{P}})}0^{(\text{R},c)}_{\text{D}}}\Big(\hat{a}_{\text{B}(N_{\text{R}},N_{\text{P}})}\Big)^{N_{-}-m-n}\Big(\hat{a}_{\text{B}(N_{\text{R}}-1,N_{\text{P}}+1)}^{\dagger}\Big)^{(\nu_++\nu_-)-(m+n)}\Big(\hat{a}^{(\text{P},c)\prime\dagger}_{\text{D}}\Big)^{\nu_{\text{D}}}\ket{0_{\text{B}(N_{\text{R}}-1,N_{\text{P}}+1)}0^{(\text{P},c)\prime}_{\text{D}}}
\end{aligned}
\end{equation}
Rewriting the initial bright mode $\hat{a}_{\text{B}(N_{\text{R}},N_{\text{P}})}$, final bright mode $\hat{a}_{\text{B}(N_{\text{R}}-1,N_{\text{P}}+1)}$, and the highly localized final dark mode $\hat{a}^{(\text{P},c)\prime}_{\text{D}}$ in terms of a bright mode involving all molecules other than the reacting molecule $\hat{a}_{\text{B}(N_{\text{R}}-1,N_{\text{P}})}=\frac{1}{\sqrt{g_{\text{R}}^2(N_{\text{R}}-1)+g_{\text{P}}^2N_{\text{P}}}}\Big[g_{\text{R}}\sum_{i=1}^{N_{\text{R}}-1}a_{\text{R},i}+g_{\text{P}}\sum_{j=1}^{N_{\text{P}}}a_{\text{P},j}\Big]$, and vibrational modes of this molecule $\hat{a}_{\text{P},c}$ and $\hat{a}_{\text{R},c}$ we obtain
\begin{equation}
\begin{aligned}
F_{\nu_+,\nu_-,\nu_{\text{D}}}^{f,0}=&\frac{1}{\sqrt{N_{-}!\nu_+!\nu_-!\nu_{\text{D}}!}}\sum_{m=0}^{\nu_+}\sum_{n=0}^{\nu_-}\binom{N_{-}}{m+n}\binom{\nu_+}{m}\binom{\nu_-}{n}\Big(\sin\theta\Big)^{m+n}\Big(\cos\theta'\Big)^m\Big(\sin\theta'\Big)^n\\
&\times\Big(-\cos\theta\Big)^{N_{-}-m-n}\Big(\sin\theta'\Big)^{\nu_+-m}\Big(-\cos\theta'\Big)^{\nu_--n}(m+n)!\\
&\times\bra{0_{\text{B}(N_{\text{R}}-1,N_{\text{P}})}0_{\text{R},c}}\Bigg[\sqrt{\frac{g_{\text{R}}^2(N_{\text{R}}-1)+g_{\text{P}}^2N_{\text{P}}}{g_{\text{R}}^2N_{\text{R}}+g_{\text{P}}^2N_{\text{P}}}}\hat{a}_{\text{B}(N_{\text{R}}-1,N_{\text{P}})}+\frac{g_{\text{R}}}{\sqrt{g_{\text{R}}^2N_{\text{R}}+g_{\text{P}}^2N_{\text{P}}}}\hat{a}_{\text{R},c}\Bigg]^{N_{-}-m-n}\\
&\times\Bigg[\sqrt{\frac{g_{\text{R}}^2(N_{\text{R}}-1)+g_{\text{P}}^2N_{\text{P}}}{g_{\text{R}}^2(N_{\text{R}}-1)+g_{\text{P}}^2(N_{\text{P}}+1)}}\hat{a}_{\text{B}(N_{\text{R}}-1,N_{\text{P}})}^{\dagger}+\frac{g_{\text{P}}}{\sqrt{g_{\text{R}}^2(N_{\text{R}}-1)+g_{\text{P}}^2(N_{\text{P}}+1)}}\hat{a}_{\text{P},c}^{\dagger}\Bigg]^{(\nu_++\nu_-)-(m+n)}\\
&\times\Bigg[-\frac{g_{\text{P}}}{\sqrt{g_{\text{R}}^2(N_{\text{R}}-1)+g_{\text{P}}^2(N_{\text{P}}+1)}}\hat{a}_{\text{B}(N_{\text{R}}-1,N_{\text{P}})}^{\dagger}+\sqrt{\frac{g_{\text{R}}^2(N_{\text{R}}-1)+g_{\text{P}}^2N_{\text{P}}}{g_{\text{R}}^2(N_{\text{R}}-1)+g_{\text{P}}^2(N_{\text{P}}+1)}}\hat{a}_{\text{P},c}^{\dagger}\Bigg]^{\nu_{\text{D}}}\ket{0_{\text{B}(N_{\text{R}}-1,N_{\text{P}})}0_{\text{P},c}}.
\end{aligned}
\end{equation}
Since $\comm{\hat{a}_{\text{B}(N_{\text{R}}-1,N_{\text{P}})}}{\hat{a}_{\text{R},c}}=0$ and $\comm{\hat{a}_{\text{B}(N_{\text{R}}-1,N_{\text{P}})}}{\hat{a}_{\text{P},c}}=0$, we can use the binomial expansion again
\begin{equation}
\begin{aligned}
F_{\nu_+,\nu_-,\nu_{\text{D}}}^{f,0}=&\frac{1}{\sqrt{N_{-}!\nu_+!\nu_-!\nu_{\text{D}}!}}\sum_{m=0}^{\nu_+}\sum_{n=0}^{\nu_-}\sum_{r=0}^{N_{-}-m-n}\sum_{p=0}^{(\nu_++\nu_-)-(m+n)}\sum_{q=0}^{\nu_{\text{D}}}\binom{N_{-}}{m+n}\binom{\nu_+}{m}\binom{\nu_-}{n}\binom{N_{-}-m-n}{r}\binom{(\nu_++\nu_-)-(m+n)}{p}\\
&\times\binom{\nu_{\text{D}}}{q}\Big(\sin\theta\Big)^{m+n}\Big(\cos\theta'\Big)^m\Big(\sin\theta'\Big)^n(m+n)!\Big(-\cos\theta\Big)^{N_{-}-m-n}\Big(\sin\theta'\Big)^{\nu_+-m}\Big(-\cos\theta'\Big)^{\nu_--n}(-1)^{q}\\
&\times
\Bigg[\frac{g_{\text{P}}}{\sqrt{g_{\text{R}}^2(N_{\text{R}}-1)+g_{\text{P}}^2(N_{\text{P}}+1)}}\Bigg]^{(\nu_++\nu_-+q)-(m+n+p)}\Bigg[\sqrt{\frac{g_{\text{R}}^2(N_{\text{R}}-1)+g_{\text{P}}^2N_{\text{P}}}{g_{\text{R}}^2(N_{\text{R}}-1)+g_{\text{P}}^2(N_{\text{P}}+1)}}\Bigg]^{\nu_{\text{D}}+p-q}\\
&\times\Bigg[\sqrt{\frac{g_{\text{R}}^2(N_{\text{R}}-1)+g_{\text{P}}^2N_{\text{P}}}{g_{\text{R}}^2N_{\text{R}}+g_{\text{P}}^2N_{\text{P}}}}\Bigg]^r\Bigg[\frac{g_{\text{R}}}{\sqrt{g_{\text{R}}^2N_{\text{R}}+g_{\text{P}}^2N_{\text{P}}}}\Bigg]^{N_{-}-m-n-r}\bra{0_{\text{B}(N_{\text{R}}-1,N_{\text{P}})}0_{\text{R},c}}\Big(\hat{a}_{\text{B}(N_{\text{R}}-1,N_{\text{P}})}\Big)^r\Big(\hat{a}_{\text{R},c}\Big)^{N_{-}-m-n-r}\\
&\times\Big(\hat{a}_{\text{B}(N_{\text{R}}-1,N_{\text{P}})}^{\dagger}\Big)^{p+q}\Big(\hat{a}_{\text{P},c}^{\dagger}\Big)^{(\nu_++\nu_-+\nu_{\text{D}})-(m+n+p+q)}\ket{0_{\text{B}(N_{\text{R}}-1,N_{\text{P}})}0_{\text{P},c}}.
\end{aligned}
\end{equation}
The vibrational modes of all molecules other than the reacting molecule are not modified by the chemical reaction, therefore, non-zero terms in the summation satisfy $r=p+q$. Plugging in $r=p+q$ and $\bra{0_{\text{B}(N_{\text{R}}-1,N_{\text{P}})}}(\hat{a}_{\text{B}(N_{\text{R}}-1,N_{\text{P}})})^{r}(\hat{a}_{\text{B}(N_{\text{R}}-1,N_{\text{P}})}^{\dagger})^{p+q}\ket{0_{\text{B}(N_{\text{R}}-1,N_{\text{P}})}}=(p+q)!$, 

\begin{equation}
\begin{aligned}
F_{\nu_+,\nu_-,\nu_{\text{D}}}^{f,0}=&\frac{1}{\sqrt{N_{-}!\nu_+!\nu_-!\nu_{\text{D}}!}}\sum_{m=0}^{\nu_+}\sum_{n=0}^{\nu_-}\sum_{p=0}^{(\nu_++\nu_-)-(m+n)}\sum_{q=0}^{\nu_{\text{D}}}\binom{N_{-}}{m+n}\binom{\nu_+}{m}\binom{\nu_-}{n}\binom{N_{-}-(m+n)}{p+q}\binom{\nu_{\text{D}}}{q}\binom{(\nu_++\nu_-)-(m+n)}{p}\\
&\times\Big(\sin\theta\Big)^{m+n}\Big(\cos\theta'\Big)^m\Big(\sin\theta'\Big)^n(m+n)!(p+q)!\Big(-\cos\theta\Big)^{N_{-}-(m+n)}\Big(\sin\theta'\Big)^{\nu_+-m}\Big(-\cos\theta'\Big)^{\nu_--n}(-1)^{q}\\
&\times\Bigg[\frac{g_{\text{P}}}{\sqrt{g_{\text{R}}^2(N_{\text{R}}-1)+g_{\text{P}}^2(N_{\text{P}}+1)}}\Bigg]^{(\nu_++\nu_-+q)-(m+n+p)}\Bigg[\sqrt{\frac{g_{\text{R}}^2(N_{\text{R}}-1)+g_{\text{P}}^2N_{\text{P}}}{g_{\text{R}}^2(N_{\text{R}}-1)+g_{\text{P}}^2(N_{\text{P}}+1)}}\Bigg]^{\nu_{\text{D}}+p-q}\Bigg[\sqrt{\frac{g_{\text{R}}^2(N_{\text{R}}-1)+g_{\text{P}}^2N_{\text{P}}}{g_{\text{R}}^2N_{\text{R}}+g_{\text{P}}^2N_{\text{P}}}}\Bigg]^{p+q}\\
&\times\Bigg[\frac{g_{\text{R}}}{\sqrt{g_{\text{R}}^2N_{\text{R}}+g_{\text{P}}^2N_{\text{P}}}}\Bigg]^{N_{-}-(m+n+p+q)}\bra{0_{\text{R},c}}\Big(\hat{a}_{\text{R},c}\Big)^{N_{-}-(m+n+p+q)}\Big(\hat{a}_{\text{P},c}^{\dagger}\Big)^{(\nu_++\nu_-+\nu_{\text{D}})-(m+n+p+q)}\ket{0_{\text{P},c}}
\end{aligned}
\end{equation}
Changing variables from $\{m, n, q, p\}$ to $\{i,j,k,h\}$, where $i=\nu_+-m$, $j=\nu_--n$, $k=\nu_{\text{D}}-q$, $h=N_{-}-m-n-p-q$ and $f=\nu_++\nu_-+\nu_{\text{D}}-N_{-}$,
\begin{equation}
\begin{aligned}
F_{\nu_+,\nu_-,\nu_{\text{D}}}^{f,0}=&\sqrt{\frac{N_{-}!}{\nu_+!\nu_-!\nu_{\text{D}}!}}\Bigg(-\cos\theta\frac{g_{\text{R}}^2(N_{\text{R}}-1)+g_{\text{P}}^2N_{\text{P}}}{g_{\text{P}}\sqrt{g_{\text{R}}^2N_{\text{R}}+g_{\text{P}}^2N_{\text{P}}}}\Bigg)^{-f}\sum_{i=0}^{\nu_+}\sum_{j=0}^{\nu_-}\sum_{k=0}^{\nu_{\text{D}}}\sum_{h=k-f}^{i+j+k-f}\binom{\nu_+}{i}\binom{\nu_-}{j}\binom{i+j}{h+f-k}\binom{\nu_{\text{D}}}{k}\\
&\times\Bigg(\frac{g_{\text{R}}g_{\text{P}}}{g_{\text{R}}^2(N_{\text{R}}-1)+g_{\text{P}}^2N_{\text{P}}}\Bigg)^h\Big(\sin\theta \cos\theta'\Big)^{\nu_+-i}\Bigg(\frac{-\cos\theta\sin\theta'(g_{\text{R}}^2(N_{\text{R}}-1)+g_{\text{P}}^2N_{\text{P}})}{\sqrt{g_{\text{R}}^2N_{\text{R}}+g_{\text{P}}^2N_{\text{P}}}\sqrt{g_{\text{R}}^2(N_{\text{R}}-1)+g_{\text{P}}^2(N_{\text{P}}+1)}}\Bigg)^{i}\Big(\sin\theta \sin\theta'\Big)^{\nu_--j}\\
&\times\Bigg(\frac{\cos\theta \cos\theta'(g_{\text{R}}^2(N_{\text{R}}-1)+g_{\text{P}}^2N_{\text{P}})}{\sqrt{g_{\text{R}}^2N_{\text{R}}+g_{\text{P}}^2N_{\text{P}}}\sqrt{g_{\text{R}}^2(N_{\text{R}}-1)+g_{\text{P}}^2(N_{\text{P}}+1)}}\Bigg)^{j}\Bigg(g_{\text{P}}\cos\theta\sqrt{\frac{g_{\text{R}}^2(N_{\text{R}}-1)+g_{\text{P}}^2N_{\text{P}}}{(g_{\text{R}}^2(N_{\text{R}}-1)+g_{\text{P}}^2(N_{\text{P}}+1))(g_{\text{R}}^2N_{\text{R}}+g_{\text{P}}^2N_{\text{P}})}}\Bigg)^{\nu_{\text{D}}-k}\\
&\times\Bigg(-\cos\theta\frac{g_{\text{R}}^2(N_{\text{R}}-1)+g_{\text{P}}^2N_{\text{P}}}{g_{\text{P}}\sqrt{g_{\text{R}}^2N_{\text{R}}+g_{\text{P}}^2N_{\text{P}}}}\sqrt{\frac{g_{\text{R}}^2(N_{\text{R}}-1)+g_{\text{P}}^2N_{\text{P}}}{g_{\text{R}}^2(N_{\text{R}}-1)+g_{\text{P}}^2(N_{\text{P}}+1)}}\Bigg)^{k}\frac{1}{h!}\bra{0_{\text{R},c}}\Big(\hat{a}_{\text{R},c}\Big)^{h}\Big(\hat{a}_{\text{P},c}^{\dagger}\Big)^{h+f}\ket{0_{\text{P},c}}
\end{aligned}
\end{equation}
The above expression is valid only when $g_{\text{P}}\neq 0$. The case of $g_{\text{P}}=0$ is explained in Subsection \textit{1}. Remembering the definitions of variables $x,y, z, w$ and $\tilde{x}, \tilde{y}, \tilde{z}$ from equation (\ref{eq:xyz}),
\begin{equation}
\begin{aligned}
F_{\nu_+,\nu_-,\nu_{\text{D}}}^{f,0}=&\sqrt{\frac{N_{-}!}{\nu_+!\nu_-!\nu_{\text{D}}!}}\tilde{w}^{-f}
\sum_{i=0}^{\nu_+}\sum_{j=0}^{\nu_-}\sum_{k=0}^{\nu_{\text{D}}}\sum_{h=k-f}^{i+j+k-f}\binom{\nu_+}{i}\binom{\nu_-}{j}\binom{i+j}{h+f-k}\binom{\nu_{\text{D}}}{k}x^{\nu_+-i}y^{\nu_--j}z^{\nu_{\text{D}}-k}\tilde{x}^i\tilde{y}^j\tilde{z}^kw^h\\
&\times\sqrt{\frac{(h+f)!}{h!}}\bra{h_{\text{R},c}}\ket{(h+f)_{\text{P},c}}
\end{aligned}
\end{equation}
This is not easy to evaluate because of the summation over $h$ involves the term $\bra{h_{\text{R},c}}\ket{(h+f)_{\text{P},c}}$. Substituting $\bra{h_{\text{R},c}}\ket{(h+f)_{\text{P},c}}=\sqrt{\frac{(h+f)!}{h!}}\sqrt{\frac{e^{-S}}{S^f}}\hat{O}_f(S)\sum_{u=0}^h\binom{h}{u}\frac{(-S)^u}{u!}$,
\begin{equation}
\begin{aligned}
F_{\nu_+,\nu_-,\nu_{\text{D}}}^{f,0}=&\sqrt{\frac{N_{-}!}{\nu_+!\nu_-!\nu_{\text{D}}!}}\tilde{w}^{-f}\sum_{i=0}^{\nu_+}\sum_{j=0}^{\nu_-}\sum_{k=0}^{\nu_{\text{D}}}\sum_{h=k-f}^{i+j+k-f}\binom{\nu_+}{i}\binom{\nu_-}{j}\binom{i+j}{h+f-k}\binom{\nu_{\text{D}}}{k}x^{\nu_+-i}y^{\nu_--j}z^{\nu_{\text{D}}-k}\tilde{x}^i\tilde{y}^j\tilde{z}^kw^h\\
&\times\frac{(h+f)!}{h!}\sqrt{\frac{e^{-S}}{S^f}}\hat{O}_f(S)\sum_{u=0}^h\binom{h}{u}\frac{(-S)^u}{u!}.
\end{aligned}
\end{equation}
Rewriting $(h+f)!w^h/h!$ as $\hat{T}_f(w)w^h$,
\begin{equation}
\begin{aligned}
F_{\nu_+,\nu_-,\nu_{\text{D}}}^{f,0}=&\sqrt{\frac{e^{-S}}{S^f}}\hat{O}_f(S)\sqrt{\frac{N_{-}!}{\nu_+!\nu_-!\nu_{\text{D}}!}}\tilde{w}^{-f}\sum_{i=0}^{\nu_+}\sum_{j=0}^{\nu_-}\sum_{k=0}^{\nu_{\text{D}}}\sum_{h=k-f}^{i+j+k-f}\sum_{u=0}^h\binom{\nu_+}{i}\binom{\nu_-}{j}\binom{i+j}{h+f-k}\binom{\nu_{\text{D}}}{k}\binom{h}{u}\\
&\times\frac{(-S)^u}{u!}x^{\nu_+-i}y^{\nu_--j}z^{\nu_{\text{D}}-k}\tilde{x}^i\tilde{y}^j\tilde{z}^k\hat{T}_f(w)w^h.
\end{aligned}
\end{equation}
With the operators $\hat{O}_f(S)$ and $\hat{T}_f(w)$ defined as:
\begin{equation}
\begin{aligned}
\hat{O}_f(S)&=
\begin{cases}
\Big(\int dS\Big)^f \hspace{30pt}&f\ge0\\
\Big(\frac{d}{dS}\Big)^{-f} \hspace{30pt}&f<0
\end{cases}
\\
\hat{T}_f(w)&=
\begin{cases}
\Big(\frac{d}{dw}\Big)^fw^f\hspace{30pt}&f\ge0\\
\Big(\int dw\Big)^{-f}w^f\hspace{30pt}&f<0
\end{cases}
\end{aligned}
\end{equation}
This simplifies to
\begin{equation}
\begin{aligned}
F^{f,0}_{\nu_+,\nu_-,\nu_{\text{D}}}
&=\sqrt{\frac{e^{-S}}{S^f}}\sqrt{\frac{N_{-}!}{\nu_+!\nu_-!\nu_{\text{D}}!}}\tilde{w}^{-f}\hat{O}_f(S)\hat{T}_f(w)\\
&\hspace{20pt}\sum_{i=0}^{\nu_+}\sum_{j=0}^{\nu_-}\sum_{k=0}^{\nu_{\text{D}}}\sum_{h=k-f}^{i+j+k-f}\sum_{u=0}^h\binom{\nu_+}{i}\binom{\nu_-}{j}\binom{i+j}{h+f-k}\binom{\nu_{\text{D}}}{k}\binom{h}{u}\frac{(-S)^u}{u!}x^{\nu_+-i}y^{\nu_--j}z^{\nu_{\text{D}}-k}\tilde{x}^i\tilde{y}^j\tilde{z}^kw^h,\\
&=\sqrt{\frac{e^{-S}}{S^f}}\sqrt{\frac{N_{-}!}{\nu_+!\nu_-!\nu_{\text{D}}!}}\tilde{w}^{-f}\hat{O}_f(S)\hat{T}_f(w)M^f_{\nu_+,\nu_-,\nu_{\text{D}}}.
\end{aligned}
\end{equation}
Here we introduce $M^f_{\nu_+,\nu_-,\nu_{\text{D}}}$ as the part involving the summation in $F^{f,0}_{\nu_+,\nu_-,\nu_{\text{D}}}$ for more readability. Evaluating $M^f_{\nu_+,\nu_-,\nu_{\text{D}}}$,
\begin{equation}
\begin{aligned}
M^f_{\nu_+,\nu_-,\nu_{\text{D}}}
=&\sum_{i=0}^{\nu_+}\sum_{j=0}^{\nu_-}\sum_{k=0}^{\nu_{\text{D}}}\sum_{h=k-f}^{i+j+k-f}\sum_{u=0}^h\binom{\nu_+}{i}\binom{\nu_-}{j}\binom{i+j}{i+j+k-f-h}\binom{\nu_{\text{D}}}{k}\binom{h}{u}\frac{(-S)^u}{u!}x^{\nu_+-i}y^{\nu_--j}z^{\nu_{\text{D}}-k}\tilde{x}^i\tilde{y}^j\tilde{z}^kw^h\\
=&\sum_{i=0}^{\nu_+}\binom{\nu_+}{i}\tilde{x}^ix^{\nu_+-i}\sum_{j=0}^{\nu_-}\binom{\nu_-}{j}\tilde{y}^jy^{\nu_--j}\sum_{k=0}^{\nu_{\text{D}}}\binom{\nu_{\text{D}}}{k}\tilde{z}^kz^{\nu_{\text{D}}-k}\sum_{h=k-f}^{i+j+k-f}\binom{i+j}{i+j+k-f-h}w^h\sum_{u=0}^h\binom{h}{u}\frac{(-S)^u}{u!}
\end{aligned}
\end{equation}
We first focus on evaluating the summation over $u$. Without $u!$ in the denominator, this would have simply been a binomial expansion. The $u!$ makes it difficult to evaluate, but each term in the summation looks like the product of terms from a binomial $\binom{h}{u}$ and exponential $(-S)^u/u!$ expansion. We can use the powerful combinatorial technique of generating functions to evaluate the summation. Here, $t$ is a dummy variable that we introduce to count the terms,
\begin{equation}
\begin{aligned}
M^f_{\nu_+,\nu_-,\nu_{\text{D}}}=&\sum_{i=0}^{\nu_+}\binom{\nu_+}{i}\tilde{x}^ix^{\nu_+-i}\sum_{j=0}^{\nu_-}\binom{\nu_-}{j}\tilde{y}^jy^{\nu_--j}\sum_{k=0}^{\nu_{\text{D}}}\binom{\nu_{\text{D}}}{k}\tilde{z}^kz^{\nu_{\text{D}}-k}\sum_{h=k-f}^{i+j+k-f}\binom{i+j}{i+j+k-f-h}w^h([t^h](1+t)^he^{-St})
\end{aligned}
\end{equation}
Each term in the summation over $h$ is a product of coefficients of two generating functions,
\begin{equation}
\begin{aligned}
M^f_{\nu_+,\nu_-,\nu_{\text{D}}}=&\sum_{i=0}^{\nu_+}\binom{\nu_+}{i}\tilde{x}^ix^{\nu_+-i}\sum_{j=0}^{\nu_-}\binom{\nu_-}{j}\tilde{y}^jy^{\nu_--j}\sum_{k=0}^{\nu_{\text{D}}}\binom{\nu_{\text{D}}}{k}\tilde{z}^kz^{\nu_{\text{D}}-k}w^{k-f}\sum_{h=k-f}^{i+j+k-f}\binom{i+j}{i+j+k-f-h}w^{h+f-k}([t^h](1+t)^he^{-St})\\
=&\sum_{i=0}^{\nu_+}\binom{\nu_+}{i}\tilde{x}^ix^{\nu_+-i}\sum_{j=0}^{\nu_-}\binom{\nu_-}{j}\tilde{y}^jy^{\nu_--j}\sum_{k=0}^{\nu_{\text{D}}}\binom{\nu_{\text{D}}}{k}\tilde{z}^kz^{\nu_{\text{D}}-k}w^{k-f}\sum_{h=k-f}^{i+j+k-f}([t^{i+j+k-f-h}](w+t)^{i+j})([t^h](1+t)^he^{-St}).
\end{aligned}
\end{equation}
For notational convenience, let us define 
\begin{equation}
\begin{aligned}
c_{\alpha}&=[t^{\alpha}](w+t)^{i+j},\hspace{1cm}
d_{\beta}&=[t^{\beta}](1+t)^{\beta}e^{-St}.
\end{aligned}
\end{equation}
Since $c_{\alpha}$ is zero when $\alpha>i+j$, the lower limit of the summation over $h$ can be shifted from $h=k-f$ to $h=0$,
\begin{equation}
\begin{aligned}
\sum_{h=k-f}^{i+j+k-f}c_{i+j+k-f-h}d_{h}=\sum_{h=0}^{i+j+k-f}c_{i+j+k-f-h}d_{h}.
\end{aligned}
\end{equation}
This summation over $h$ can be written as the coefficient of the product of two generating function 
\begin{equation}
\begin{aligned}
\sum_{h=0}^{i+j+k-f}c_{i+j+k-f-h}d_{h}=
[q^{i+j+k-f}]\bigg(\sum_{\alpha=0}^{\infty}c_{\alpha}q^{\alpha}\bigg)
\bigg(\sum_{\beta=0}^{\infty}d_{\beta}q^{\beta}\bigg),
\end{aligned}
\end{equation}
where $q$ is the new dummy variable. Now we will find a closed expression for the generating function
\begin{equation}
\begin{aligned}
C(q)&=\sum_{\alpha=0}^{\infty}c_{\alpha}q^{\alpha}, \hspace{1cm}
D(q)&=\sum_{\beta=0}^{\infty}d_{\beta}q^{\beta}.
\end{aligned}
\end{equation} 
Using binomial expansion, we get the closed expression $C(q)=(w+q)^{i+j}$. However, it is much more complicated to obtain a closed expression for $D(q)$. To do so, we use the powerful  Lagrange-B\"urmann formula, which states that for any generating series $f(t)$ and $g(t)$ such that $f(0)\ne 0$ and a change of variable $q=t/f(t)$, we have an identity of generating functions in the dummy variable $q$,
\begin{equation}
\sum_{\beta=0}^{\infty}([t^{\beta}]f(t)^{\beta}g(t))q^{\beta}=\frac{g(t)}{f(t)}\frac{dt}{dq},
\end{equation}
where $t$ is expressed in terms of $q$. In our scenario, $f(t)=(1+t)$ and $g(t)=e^{-St}$, hence $q=t/(1+t)$, which gives us
\begin{equation}
\begin{aligned}
t=\frac{q}{1-q}\hspace{1cm}
\frac{dt}{dq}=\frac{1}{(1-q)^2}
\end{aligned}
\end{equation}
Writing $t$ in terms of $q$ we get $f(t)
=\frac{1}{1-q}$ and $g(t)=e^{-Sq/(1-q)}$. Therefore we have 
\begin{equation}
\begin{aligned}
D(q)&=\frac{e^{-Sq/(1-q)}}{(1-q)^2}(1-q)\\
&=\frac{e^{-Sq/(1-q)}}{(1-q)}.
\end{aligned}
\end{equation}
Continuing with the original summation, we now have
\begin{equation}
\begin{aligned}
M^f_{\nu_+,\nu_-,\nu_{\text{D}}}&=\sum_{i=0}^{\nu_+}\binom{\nu_+}{i}\tilde{x}^ix^{\nu_+-i}\sum_{j=0}^{\nu_-}\binom{\nu_-}{j}\tilde{y}^jy^{\nu_--j}\sum_{k=0}^{\nu_{\text{D}}}\binom{\nu_{\text{D}}}{k}\tilde{z}^kz^{\nu_{\text{D}}-k}w^{k-f}\Big([t^{i+j+k-f}](w+t)^{i+j}\frac{e^{-St/(1-t)}}{(1-t)}\Big)
\end{aligned}
\end{equation}
Note that in the above expression we again replaced $q$ with $t$. Repeating the calculations using Lagrange-B\"urmann formula three more times, for the summation over $i,j$ and $k$, we get the final compressed expression of the desired summation 
\begin{equation}
M^f_{\nu_+,\nu_-,\nu_{\text{D}}}
=[q^{N_{-}}]\Bigg((\tilde{x}+qx)^{\nu_+}(\tilde{y}+yq)^{\nu_-}\frac{(\tilde{z}-\tilde{z}q+zq)^{\nu_{\text{D}}}}{(1-q)^{f}}\frac{\exp(\frac{-Swq}{1-q-wq})}{(1-q-wq)}\Bigg).
\end{equation}
Applying operators $\hat{O}_f(S)$ and $\hat{T}_f(w)$ to $M^f_{\nu_+,\nu_-,\nu_{\text{D}}}$
\begin{equation}\label{eq:gPneq0}
\begin{aligned}
F^{f,0}_{\nu_+,\nu_-,\nu_{\text{D}}}&=\sqrt{\frac{e^{-S}}{S^f}}\sqrt{\frac{N_{-}!}{\nu_+!\nu_-!\nu_{\text{D}}!}}\tilde{w}^{-f}[q^{N_{-}}]\Bigg((\tilde{x}+qx)^{\nu_+}(\tilde{y}+yq)^{\nu_-}\frac{(\tilde{z}-\tilde{z}q+zq)^{\nu_{\text{D}}}}{(1-q)^{f}}\hat{O}_f(S)\hat{T}_f(w)\frac{\exp(\frac{-Swq}{1-q-wq})}{(1-q-wq)}\Bigg)\\
&=\sqrt{e^{-S}S^f}\sqrt{\frac{N_{-}!}{\nu_+!\nu_-!\nu_{\text{D}}!}}\tilde{w}^{-f}[q^{N_{-}}]\Bigg((\tilde{x}+qx)^{\nu_+}(\tilde{y}+yq)^{\nu_-}(\tilde{z}-\tilde{z}q+zq)^{\nu_{\text{D}}}\frac{\exp(\frac{-Swq}{1-q-wq})}{(1-q-wq)^{f+1}}\Bigg)
\end{aligned}
\end{equation}

\subsubsection{Product not coupled}
When $g_{\text{P}}= 0$, the expression in equation (\ref{eq:gPneq0}) does not apply,
\begin{equation}
\begin{aligned}
F_{\nu_+,\nu_-,\nu_{\text{D}}}^{f,0}=&\sqrt{\frac{N_{-}!}{\nu_+!\nu_-!\nu_{\text{D}}!}}\sum_{m=0}^{\nu_+}\sum_{n=0}^{\nu_-}\sum_{q=0}^{\nu_{\text{D}}}\sum_{h=\nu_{\text{D}}-f-q}^{N_{-}-m-n-q}\binom{\nu_+}{m}\binom{\nu_-}{n}\binom{(\nu_++\nu_-)-(m+n)}{h+q+f-\nu_{\text{D}}}\binom{\nu_{\text{D}}}{q}\\
&\times\Big(\sin\theta\Big)^{m+n}\Big(\cos\theta'\Big)^m\Big(\sin\theta'\Big)^n\Big(-\cos\theta\Big)^{N_{-}-(m+n)}\Big(\sin\theta'\Big)^{\nu_+-m}\Big(-\cos\theta'\Big)^{\nu_--n}(-1)^{q}\\
&\times\Bigg[\frac{g_{\text{P}}}{\sqrt{g_{\text{R}}^2(N_{\text{R}}-1)+g_{\text{P}}^2(N_{\text{P}}+1)}}\Bigg]^{(f+h+2q)-\nu_{\text{D}}}\Bigg[\sqrt{\frac{g_{\text{R}}^2(N_{\text{R}}-1)+g_{\text{P}}^2N_{\text{P}}}{g_{\text{R}}^2(N_{\text{R}}-1)+g_{\text{P}}^2(N_{\text{P}}+1)}}\Bigg]^{\nu_{\text{D}}+N_{-}-h-m-n-2q}\\
&\times\Bigg[\sqrt{\frac{g_{\text{R}}^2(N_{\text{R}}-1)+g_{\text{P}}^2N_{\text{P}}}{g_{\text{R}}^2N_{\text{R}}+g_{\text{P}}^2N_{\text{P}}}}\Bigg]^{N_{-}-h-m-n}\Bigg[\frac{g_{\text{R}}}{\sqrt{g_{\text{R}}^2N_{\text{R}}+g_{\text{P}}^2N_{\text{P}}}}\Bigg]^{h}\sqrt{\frac{(h+f)!}{h!}}\bra{h_{\text{R},c}}\ket{(h+f)_{\text{P},c}}.
\end{aligned}
\end{equation}
Since $g_{\text{P}}=0$, only terms with $f+h+2q-\nu_{\text{D}}=0$ will be non-zero,
\begin{equation}
\begin{aligned}
F_{\nu_+,\nu_-,\nu_{\text{D}}}^{f,0}=&\sqrt{\frac{N_{-}!}{\nu_+!\nu_-!\nu_{\text{D}}!}}\sum_{m=0}^{\nu_+}\sum_{n=0}^{\nu_-}\sum_{q=0}^{\nu_{\text{D}}}\binom{\nu_+}{m}\binom{\nu_-}{n}\binom{(\nu_++\nu_-)-(m+n)}{-q}\binom{\nu_{\text{D}}}{q}\Big(\sin\theta\Big)^{m+n}\Big(\cos\theta'\Big)^m\\
&\times\Big(\sin\theta'\Big)^n\Big(-\cos\theta\Big)^{N_{-}-(m+n)}\times\Big(\sin\theta'\Big)^{\nu_+-m}\Big(-\cos\theta'\Big)^{\nu_--n}(-1)^{q}\Bigg(\frac{1}{\sqrt{N_{\text{R}}}}\Bigg)^{\nu_{\text{D}}-f-2q}\\
&\times\Bigg(\sqrt{\frac{N_{\text{R}}-1}{N_{\text{R}}}}\Bigg)^{N_{-}+f+2q-\nu_{\text{D}}-m-n}\sqrt{\frac{(\nu_{\text{D}}-2q)!}{(\nu_{\text{D}}-2q-f)!}}\bra{(\nu_{\text{D}}-2q-f)_{\text{R},c}}\ket{(\nu_{\text{D}}-2q)_{\text{P},c}}.
\end{aligned}
\end{equation}
From the above, we see that $q=0$ because of the binomial coefficient involving $-q$ and
\begin{equation}
\begin{aligned}
F_{\nu_+,\nu_-,\nu_{\text{D}}}^{f,0}=&\sqrt{\frac{N_{-}!}{\nu_+!\nu_-!\nu_{\text{D}}!}}\sum_{m=0}^{\nu_+}\sum_{n=0}^{\nu_-}\binom{\nu_+}{m}\binom{\nu_-}{n}\Big(\sin\theta\Big)^{m+n}\Big(\cos\theta'\Big)^m\Big(\sin\theta'\Big)^n\Big(-\cos\theta\Big)^{N_{-}-(m+n)}\Big(\sin\theta'\Big)^{\nu_+-m}\\
&\times\Big(-\cos\theta'\Big)^{\nu_--n}\Bigg(\frac{1}{\sqrt{N_{\text{R}}}}\Bigg)^{\nu_{\text{D}}-f}\Bigg(\sqrt{\frac{N_{\text{R}}-1}{N_{\text{R}}}}\Bigg)^{N_{-}+f-\nu_{\text{D}}-m-n}\sqrt{\frac{\nu_{\text{D}}!}{(\nu_{\text{D}}-f)!}}\bra{(\nu_{\text{D}}-f)_{\text{R},c}}\ket{(\nu_{\text{D}})_{\text{P},c}}.
\end{aligned}
\end{equation}
Using the binomial expansion to collect terms, we get
\begin{equation}
\begin{aligned}
F_{\nu_+,\nu_-,\nu_{\text{D}}}^{f,0}=&\sqrt{\frac{N_{-}!}{\nu_+!\nu_-!(\nu_{\text{D}}-f)!}}(x+\tilde{x})^{\nu_+}(y+\tilde{y})^{\nu_-}u^{\nu_{\text{D}}-f}\bra{(\nu_{\text{D}}-f)_{\text{R},c}}\ket{(\nu_{\text{D}})_{\text{P},c}}.
\end{aligned}
\end{equation}
\subsubsection{Reactant not coupled}
Starting from equation (\ref{eq:fcfactors}) and substituting $g_{\text{R}}=0$, we have $w=0$,
\begin{equation}
\begin{aligned}
F^{f,0}_{\nu_+,\nu_-,\nu_{\text{D}}}=&\sqrt{e^{-S}S^f}\sqrt{\frac{N_{-}!}{\nu_+!\nu_-!\nu_{\text{D}}!}}\tilde{w}^{-f}\Big[q^{N_{-}}\Big]\Bigg((\tilde{x}+xq)^{\nu_+}(\tilde{y}+yq)^{\nu_-}(\tilde{z}-\tilde{z}q+zq)^{\nu_{\text{D}}}\frac{1}{(1-q)^{f+1}}\Bigg).
\end{aligned}
\end{equation}
Changing variable from $q$ to $t=q(1+w)$
\begin{equation}
\begin{aligned}
F^{f,0}_{\nu_+,\nu_-,\nu_{\text{D}}}=&\sqrt{e^{-S}S^f}\sqrt{\frac{N_{-}!}{\nu_+!\nu_-!\nu_{\text{D}}!}}(\tilde{w}(1+w))^{-f}[t^{N_{-}}]\Bigg(\Big(\tilde{x}(1+w)+xt\Big)^{\nu_+}\Big(\tilde{y}(1+w)+yt\Big)^{\nu_-}\\
&\times\Big(\tilde{z}(1+w)+(z-\tilde{z})t\Big)^{\nu_{\text{D}}}\frac{1}{(1-t)^{f+1}}\Bigg)\\
=&\sqrt{e^{-S}S^f}\sqrt{\frac{N_{-}!}{\nu_+!\nu_-!\nu_{\text{D}}!}}(\tilde{w}(1+w))^{-f}B_{\nu_+,\nu_-,\nu_{\text{D}}}^f,
\end{aligned}
\end{equation}
and we call the part of $F^{f,0}_{\nu_+,\nu_-,\nu_{\text{D}}}$ that involves taking the $t^{N_{-}}$ coefficient
 \begin{equation}
 \begin{aligned}
 B_{\nu_+,\nu_-,\nu_{\text{D}}}^f
 &=[t^{N_{-}}]\Bigg(\frac{G_{\nu_+,\nu_-,\nu_{\text{D}}}(t)}{(1-t)^{f+1}}\Bigg),
 \end{aligned}
 \end{equation}
where
\begin{equation}
\begin{aligned}
G_{\nu_+,\nu_-,\nu_{\text{D}}}(t)&=\Big(\tilde{x}(1+w)+xt\Big)^{\nu_+}\Big(\tilde{y}(1+w)+yt\Big)^{\nu_-}\Big(\tilde{z}(1+w)+(z-\tilde{z})t\Big)^{\nu_{\text{D}}}.
\end{aligned}
\end{equation}
When $f<0$, then $\deg(G_{\nu_+,\nu_-,\nu_{\text{D}}}(t)(1-t)^{-f-1})=N_{-}-1$, therefore, $B^f_{\nu_+,\nu_-,\nu_{\text{D}}}=0$. When $f=0$, then $\deg(G_{\nu_+,\nu_-,\nu_{\text{D}}}(t))=N_{-}$ and we have the following identity
\begin{equation}
[t^{N_{-}}]\Bigg(\frac{G_{\nu_+,\nu_-,\nu_{\text{D}}}(t)}{(1-t)}\Bigg)=G_{\nu_+,\nu_-,\nu_{\text{D}}}(1).
\end{equation}
Using this identity, we obtain the expression for $B^f_{\nu_+,\nu_-,\nu_{\text{D}}}$ base case $f=0$,
\begin{equation}
B^0_{\nu_+,\nu_-,\nu_{\text{D}}}=G_{\nu_+,\nu_-,\nu_{\text{D}}}(1).
\end{equation}
Now that we have the result for $f=0$, let's derive a recursive formula for $B^f_{\nu_+,\nu_-,\nu_{\text{D}}}$ when $f\ge1$. The coefficient of $t^{N_{-}+1}$ in a series $\sum_n a_nt^n$ is related to the coefficient of $t^{N_{-}}$ of the derivative of the same series. Using this,
\begin{equation}
\begin{aligned}
\left[t^{N_{-}}\right]\frac{d}{dt}\Bigg(\frac{G_{\nu_+,\nu_-,\nu_{\text{D}}}(t)}{(1-t)^{f}}\Bigg)&=(N_{-}+1)[t^{N_{-}+1}]\Bigg(\frac{G_{\nu_+,\nu_-,\nu_{\text{D}}}(t)}{(1-t)^{f}}\Bigg)\\
[t^{N_{-}}]\Bigg(\frac{G_{\nu_+,\nu_-,\nu_{\text{D}}}'(t)}{(1-t)^{f}}\Bigg)+f[t^{N_{-}}]\Bigg(\frac{G_{\nu_+,\nu_-,\nu_{\text{D}}}(t)}{(1-t)^{f+1}}\Bigg)&=(N_{-}+1)[t^{N_{-}+1}]\Bigg(\frac{G_{\nu_+,\nu_-,\nu_{\text{D}}}(t)}{(1-t)^{f}}\Bigg)\\
fB_{\nu_+,\nu_-,\nu_{\text{D}}}^f&=(N_{-}+1)B_{\nu_+,\nu_-,\nu_{\text{D}}}^{f-1}-[t^{N_{-}}]\Bigg(\frac{G_{\nu_+,\nu_-,\nu_{\text{D}}}'(t)}{(1-t)^{f}}\Bigg)
\end{aligned}
\end{equation}
\begin{equation}
\begin{aligned}
B_{\nu_+,\nu_-,\nu_{\text{D}}}^f&=\frac{1}{f}\Bigg((N_{-}+1)B_{\nu_+,\nu_-,\nu_{\text{D}}}^{f-1}-\nu_+xB_{\nu_+-1,\nu_-,\nu_{\text{D}}}^{f-1}-\nu_-yB_{\nu_+,\nu_--1,\nu_{\text{D}}}^{f-1}-\nu_{\text{D}}(z-\tilde{z})B_{\nu_+,\nu_-,\nu_{\text{D}}-1}^{f-1}\Bigg)
\end{aligned}
\end{equation}
The expression for $B_{\nu_+,\nu_-,\nu_{\text{D}}}^f$ for all the different cases,
\begin{equation}
\begin{aligned}
B^{f}_{\nu_+,\nu_-,\nu_{\text{D}}}=
\begin{cases}
0 &f<0\\
G_{\nu_+,\nu_-,\nu_{\text{D}}}(1) &f=0\\
\frac{1}{f}\Bigg((N_{-}+1)B_{\nu_+,\nu_-,\nu_{\text{D}}}^{f-1}-\nu_+xB_{\nu_+-1,\nu_-,\nu_{\text{D}}}^{f-1}-\nu_-yB_{\nu_+,\nu_--1,\nu_{\text{D}}}^{f-1}-\nu_{\text{D}}(z-\tilde{z})B_{\nu_+,\nu_-,\nu_{\text{D}}-1}^{f-1}\Bigg) & f>0
\end{cases}
\end{aligned}
\end{equation}

\subsubsection{Product and reactant equally coupled}
For the special case when $g_{\text{R}}=g_{\text{P}}$, we change variable $t=q(1+w)$ and the expression becomes
\begin{equation}
\begin{aligned}
F_{\nu_+,\nu_-,\nu_{\text{D}}}^{f,0}
=&\sqrt{e^{-S}S^f}\sqrt{\frac{N_{-}!}{\nu_+!\nu_-!\nu_{\text{D}}!}}(\tilde{w}(1+w))^{-f}[t^{N_{-}}]\Bigg(\Big(\tilde{x}(1+w)+xt\Big)^{\nu_+}\Big(\tilde{y}(1+w)+yt\Big)^{\nu_-}\\
&\times\Big(\tilde{z}(1+w)+(z-\tilde{z})t\Big)^{\nu_{\text{D}}}\frac{\exp(\frac{-Swt}{(1+w)(1-t)})}{(1-t)^{f+1}}\Bigg)\\
=&\sqrt{e^{-S}S^f}\sqrt{\frac{Q!}{\nu_+!\nu_-!\nu_{\text{D}}!}}(\tilde{w}(1+w))^{-f}A_{\nu_+,\nu_-,\nu_{\text{D}}}^{f},
\end{aligned}
\end{equation}
where
\begin{equation}
\begin{aligned}
A_{\nu_+,\nu_-,\nu_{\text{D}}}^f&=[t^{N_{-}}]\Bigg(\Big(\tilde{x}(1+w)+xt\Big)^{\nu_+}\Big(\tilde{y}(1+w)+yt\Big)^{\nu_-}\Big(\tilde{z}(1+w)+(z-\tilde{z})t\Big)^{\nu_{\text{D}}}\frac{\exp(\frac{-Swt}{(1+w)(1-t)})}{(1-t)^{f+1}}\Bigg).
\end{aligned}
\end{equation}
Expanding the exponential,
\begin{equation}
\begin{aligned}
A_{\nu_+,\nu_-,\nu_{\text{D}}}^f&=[t^Q]\Bigg(\Big(\tilde{x}(1+w)+xt\Big)^{\nu_+}\Big(\tilde{y}(1+w)+yt\Big)^{\nu_-}\Big(\tilde{z}(1+w)+(z-\tilde{z})t\Big)^{\nu_{\text{D}}}\frac{1}{(1-t)^{f+1}}exp\Big(\frac{-Swt}{(1+w)(1-t)}\Big)\Bigg)\\
&=\sum_{n=0}^{\infty}\frac{1}{n!}[t^Q]\Bigg(\Big(\tilde{x}(1+w)+xt\Big)^{\nu_+}\Big(\tilde{y}(1+w)+yt\Big)^{\nu_-}\Big(\tilde{z}(1+w)+(z-\tilde{z})t\Big)^{\nu_{\text{D}}}\frac{1}{(1-t)^{f+1}}\Big(\frac{-Swt}{(1+w)(1-t)}\Big)^n\Bigg)\\
&=\sum_{n=0}^{\infty}w^n\frac{1}{n!}\Big(\frac{-S}{1+w}\Big)^n[t^Q]\Bigg(\Big(\tilde{x}(1+w)+xt\Big)^{\nu_+}\Big(\tilde{y}(1+w)+yt\Big)^{\nu_-}\Big(\tilde{z}(1+w)+(z-\tilde{z})t\Big)^{\nu_{\text{D}}}\frac{t^n}{(1-t)^{f+1+n}}\Bigg)\\
&=\sum_{n=0}^{\infty}w^n\frac{1}{n!}\Big(\frac{-S}{1+w}\Big)^n[t^{Q-n}]\Bigg(\Big(\tilde{x}(1+w)+xt\Big)^{\nu_+}\Big(\tilde{y}(1+w)+yt\Big)^{\nu_-}\Big(\tilde{z}(1+w)+(z-\tilde{z})t\Big)^{\nu_{\text{D}}}\frac{1}{(1-t)^{f+1+n}}\Bigg)\\
&=\sum_{n=0}^{\infty}w^n\frac{1}{n!}\Big(\frac{-S}{1+w}\Big)^nB^{f+n}_{\nu_+,\nu_-,\nu_{\text{D}}}
\end{aligned}
\end{equation}
For the case when $g_{\text{R}}=g_{\text{P}}$, $w=1/(N-1)$ is very small when a large number of molecules are coupled to the cavity. Therefore, the summation in $A^f_{\nu_+,\nu_-,\nu_{\text{D}}}$ converges quickly and is easy to calculate on a computer.

\bibliography{supplementary}